\documentclass[preprint]{aastex}

\newcommand{\ltsimeq}{\raisebox{-0.6ex}{$\,\stackrel
        {\raisebox{-.2ex}{$\textstyle <$}}{\sim}\,$}}
\newcommand{\gtsimeq}{\raisebox{-0.6ex}{$\,\stackrel
        {\raisebox{-.2ex}{$\textstyle >$}}{\sim}\,$}}

\received{}
\revised{}
\accepted{}

\shortauthors{Boyer et al.}
\shorttitle{``AGB Census in Local Group Dwarf Galaxies''}

\usepackage[dvips,width=6.5in,height=9in,top=1in,left=1in]{geometry}

\begin{document}

\title{A \emph{SPITZER} STUDY OF ASYMPTOTIC GIANT BRANCH STARS. III. DUST
  PRODUCTION AND GAS RETURN IN LOCAL GROUP DWARF IRREGULAR
  GALAXIES.}

\author{Martha L. Boyer\altaffilmark{1}, Evan
  D. Skillman\altaffilmark{2}, Jacco Th. van Loon\altaffilmark{3},
  Robert D. Gehrz\altaffilmark{2}, and Charles
  E. Woodward\altaffilmark{2}}\altaffiltext{1}{STScI, 3700 San Martin
  Drive Baltimore, MD 21218 USA; mboyer@stsci.edu}
  \altaffiltext{2}{Astronomy Department, School of Physics and
  Astronomy, 116 Church Street, S.E., University of Minnesota,
  Minneapolis, MN 55455 USA} \altaffiltext{3}{Astrophysics Group,
  School of Physical \& Geographical Sciences, Keele University,
  Staffordshire ST5 5BG, UK}

\begin{abstract}
We present the third and final part of a census of Asymptotic Giant
Branch (AGB) stars in Local Group dwarf irregular galaxies. Papers~I
and II presented the results for WLM and IC~1613. Included here are
Phoenix, LGS~3, DDO~210, Leo~A, Pegasus~dIrr, and Sextans~A. {\it
Spitzer} photometry at 3.6, 4.5, 5.8, and 8~\micron{} are presented,
along with a more thorough treatment of background galaxy
contamination than was presented in papers~I and II.  We find that at
least a small population of completely optically obscured AGB stars
exists in each galaxy, regardless of the galaxy's metallicity, but
that higher-metallicity galaxies tend to harbor more stars with slight
IR excesses. The optical incompleteness increases for the redder AGB
stars, in line with the expectation that some AGB stars are not
detected in the optical due to large amounts of extinction associated
with in situ dust production. Overall, there is an underrepresentation
of 30\% -- 40\% in the optical AGB within the 1\,$\sigma$ errors for
all of the galaxies in our sample. This undetected population is large
enough to affect star formation histories derived from optical
color-magnitude diagrams. As measured from the $[3.6] - [4.5]$ color
excesses, we find average stellar mass-loss rates ranging from $3.1
\times 10^{-7}\,- \,6.6 \times 10^{-6}~M_\odot~{\rm yr}^{-1}$, and
integrated galaxy mass-loss rates ranging from $4.4 \times
10^{-5}\,-\,1.4 \times 10^{-3}~M_\odot~{\rm yr}^{-1}$. The integrated
mass-loss rate is sufficient to sustain the current star formation
rate in only LGS~3 and DDO~210, requiring either significant non-dusty
mass loss or gas accretion in Phoenix, Leo~A, Pegasus~dIrr, Sextans~A,
WLM, and IC~1613 if they are to maintain their status as gas-rich
galaxies.
\end{abstract}


\vfill\eject 
\section{INTRODUCTION}
\label{sec:intro}

After a star with mass $0.8~M_\odot < M < 8~M_\odot$ has exhausted
the helium in its core, it will experience a short stay
\citep[$\approx$1 -- 13~Myr;][]{vassiliadis93} on the Asymptotic
Giant Branch (AGB), where it will burn hydrogen and helium in
concentric shells around a degenerate C/O core. During the final
0.2 -- 2 Myr of its life, an AGB star will undergo radial pulses,
which may allow dust condensation in the circumstellar
envelope. Radiation pressure on the dust grains and momentum coupling
between the grains and the gas results in the formation of an
additional slow, dense wind \citep{gehrz71}. This wind causes the star to lose
20 -- 60\% of its mass, truncating its subsequent evolution.

AGB stars are uniquely important to many aspects of galactic and
stellar astronomy. Mass-losing AGB stars are the main source of
long-term gas return to the interstellar medium (ISM), prolonging the
star-formation life of gas-rich galaxies. AGB winds are heavily
enriched in Li, C, N, F, and s-process elements, and are the largest
source of dust input into the ISM \citep{gehrz89}, making AGB stars an
important driver of chemical evolution in galaxies. The intermediate
masses of AGB stars make them ideal candidates for probing the star
formation histories (SFHs) across much of a galaxy's lifetime.  Also,
because they are widely distributed and easily detected, AGB stars are
excellent tracers of galactic structure. While unobscured AGB stars
are among the brightest stars in optical band-passes, they are best
studied in the infrared (IR) and the near-IR, where emission from warm
circumstellar dust peaks and circumstellar extinction is
minimal. However, a lack of sufficient sensitivity and resolution has
prevented comprehensive studies in the near-IR and IR in systems other
than the Magellanic Clouds \citep[e.g.,][]{loup97, trams99,
cioni03,whitelock03}. While near-IR observations of distant stellar
systems remain difficult, the recent launch of the {\it Spitzer Space
Telescope} has overcome most IR observing obstacles.

Dwarf irregular (dIrr) galaxies in the Local Group provide us with
ideal laboratories to study AGB stars.  Most of these galaxies are
undergoing recent or on-going star formation, resulting in a
population of AGB stars representative of the entire mass range rather
than a snapshot of a single AGB mass.  The nearby dwarf irregular
galaxies are populous enough to capture a substantial number of AGB
stars in the short-lived mass-loss phase, near enough to resolve the
stellar population, and distant enough to encompass the entire galaxy
with a small field of view. {\it Spitzer's} ability to detect the
entire AGB population in a given galaxy also provides us with the
means to study the mass and dust injection rate into the Interstellar
Medium (ISM), the fate of the ISM, and the effect on continued star
formation and chemical evolution of the galaxy.

This paper is the third and final part of our {\it Spitzer} mid-IR
IRAC survey of Local Group dIrr galaxies, designed to obtain a
complete census of AGB stars.  \citet{jackson07a} (hereafter paper~I),
and \citet{jackson07b} (hereafter paper~II) presented a complete AGB
census in WLM and IC~1613, respectively.  Here, we present the results
for six more Local Group dIrr galaxies and provide a comprehensive
summary of the results for all eight galaxies. Our total sample of
eight galaxies spans 1~dex in metallicity, allowing us to study AGB
dust production and mass loss in metal-poor environments as a function
of metallicity. This metallicity range also enables us to investigate
the degree to which optical studies of these targets are biased by the
inability to detect the entire AGB population due to dust obscuration.

In \S\,\ref{sec:obs}, we describe the observations, data reduction,
and ancillary data. We present and describe the IR color-magnitude
diagrams (CMDs) in \S\,\ref{sec:compare} and also discuss the stellar
spatial distributions, detection statistics, stellar luminosity, and
stellar mass. \S\,\ref{sec:agbs} details a complete census of the AGB
stars in all eight galaxies and estimates of AGB mass-loss rates
(MLRs) and ISM gas return. We summarize our results in
\S\,\ref{sec:summary}.

\section{OBSERVATIONS AND DATA REDUCTION}
\label{sec:obs}

\subsection{\emph {Sample Selection}}
\label{sec:sample}

Papers~I and II discuss the AGB populations in the Local Group dIrr
galaxies WLM and IC~1613, which are very similar in
luminosity, gas content, and metallicity. In order to broaden the
environments for which we have complete AGB censuses, we have chosen
six more Local Group dIrr galaxies for this study: Phoenix dwarf,
LGS~3 (Pisces~dIrr), DDO~210 (Aquarius~dIrr), Leo~A (DDO~69),
Pegasus~dIrr (DDO~216), and Sextans~A (DDO~75). Including these six
galaxies with WLM and IC~1613 increases the luminosity range to
4.6~mag, the range in \ion{H}{1} mass by a factor of 800, and the
range in 12\,+\,log(O/H) to 0.9. See Table~\ref{tab:properties} for a
list of basic galaxy properties.

The angular sizes of the galaxies presented here are small enough to
obtain complete coverage, and the galaxies are near enough to resolve
stars down to or below the tip of the Red Giant Branch (TRGB), above
which mass-losing AGB stars reside. The span in metallicity \citep[$Z
\approx$ 2\% -- 19\%~$Z_\odot$;][]{lee06,vanzee06}, provides us with
the opportunity to examine how metal content affects dust production,
mass return, and optical completeness (\S\,\ref{sec:agbs}).

Each galaxy shows some evidence for recent star formation \citep[age
$<$ 3~Gyr;][]{dolphin05}, providing a large sample of
intermediate-aged stars, and consequently, a large sample of AGB
stars.  LGS~3 has the lowest rate of recent star formation in our
sample, followed closely by Phoenix. Sextans~A, WLM, and IC~1613 are
forming stars at a rate two orders of magnitude higher than LGS~3. A
brief summary of the star formation histories from \citet{dolphin05}
is given in \S\,\ref{sec:compare}.

\subsection{\emph {Infrared Data and Photometry}}
\label{sec:irphot}

\begin{figure}[h!]
\epsscale{1} \plotone{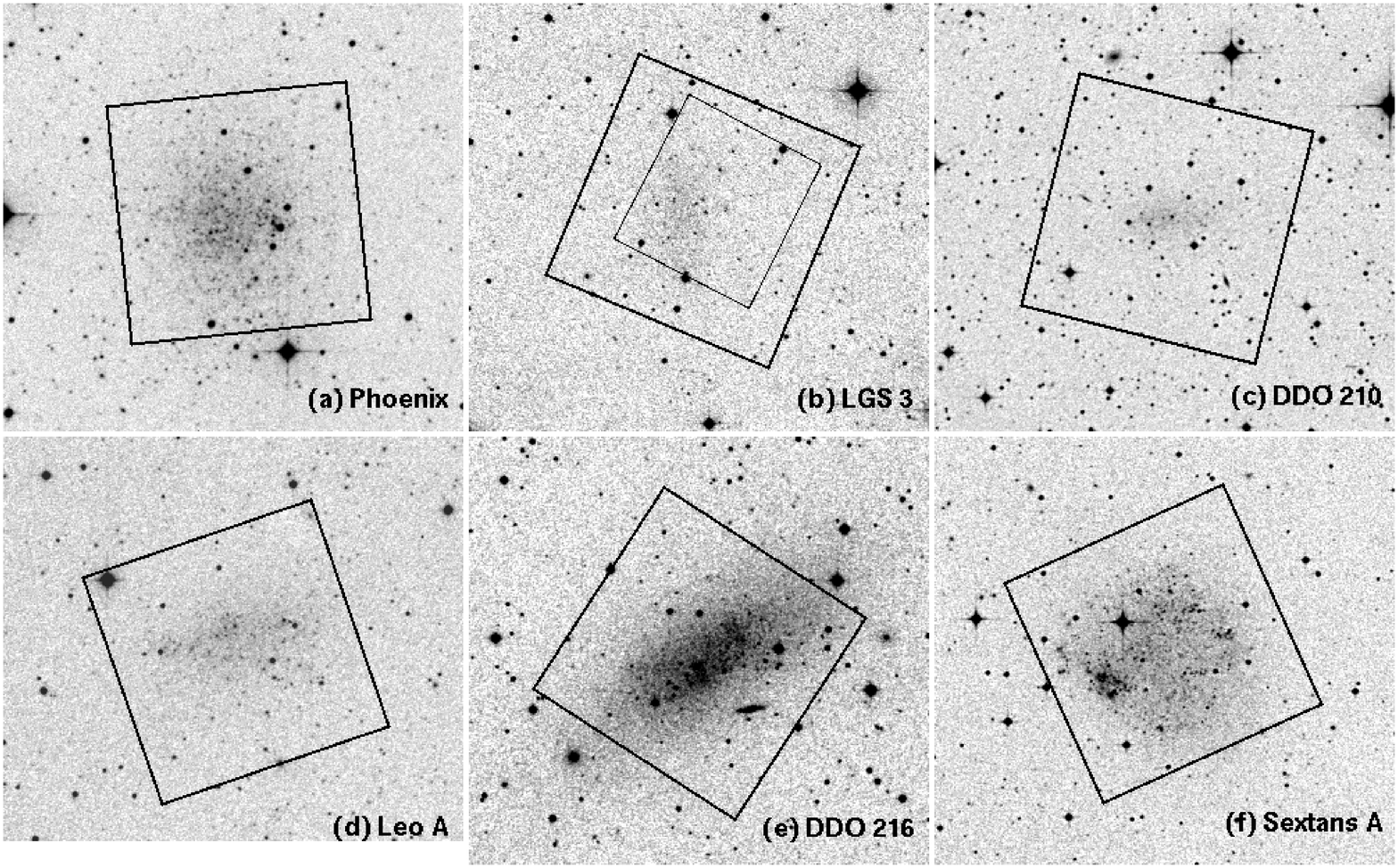} \figcaption{Digitized Sky
Survey (DSS) images of (a)~Phoenix, (b)~LGS~3, (c)~DDO~210, (d)~Leo~A,
(e)~Pegasus~dIrr, and (f)~Sextans~A. The field shown in each panel is
9.7\arcmin{} $\times$ 9.5\arcmin{}. The IRAC field of view is
overplotted with a thick line. The optical coverage for each galaxy,
with the exception of LGS~3, is larger than the DSS image shown. (b)
shows the HST ACS field of view for LGS~3 plotted as a thin black
line. In all six galaxies, the IRAC coverage contains the vast
majority of the stellar population.\label{fig:irfov} }
\end{figure}

New observations were obtained with the Infrared Array Camera
\citep[IRAC;][]{fazio04} onboard the {\it Spitzer Space Telescope}
\citep{werner04,gehrz07}. The observing program (PID 40524) was
designed to image Phoenix, LGS~3, DDO~210, Leo~A, and Pegasus~dIrr at
3.6 and 5.8~\micron{}.  The same five galaxies were also observed
earlier at 4.5 and 8~\micron{} as part of a larger guaranteed time
observing (GTO) program (PID 128; PI: R.~D.~Gehrz). Sextans~A was
observed as part of the GTO program at all four IRAC wavelengths. See
Table~\ref{tab:obs} for a summary of the observations.  GTO 4.5 and
8~\micron{} data were observed with a 5-point Gaussian dither pattern
\citep{handbook} and an exposure time of 193.6~s, yielding a total
depth of 968~s. To build a better redundancy against outliers and
artifacts while maintaining a similar exposure depth, images obtained
as part of PID 40524 were observed with a 36-point reuleaux dither
pattern \citep{handbook} and exposure time of 26.8~s, for a total
depth of 965~s. Sextans~A was observed at 3.6 and 5.8~\micron{} as
part of PID 128 with a 16-point reuleaux dither pattern and exposure
time of 26.8~s, for a total depth of 858~s.  For all observations, we
used small-scale dither patterns to allow for sub-pixel
sampling. These observations yielded 5\,$\sigma$ point-source
sensitivities of 1.3, 2.3, 15.0, and 17.5~$\mu$Jy at 3.6, 4.5, 5.8,
and 8~\micron{}, respectively. Sextans~A 5\,$\sigma$ sensitivities are
1.4~$\mu$Jy at 3.6~\micron{} and 15.9~$\mu$Jy at 5.8~\micron{}. These
sensitivities were computed using the {\it Spitzer} sensitivity
calculator\footnote{http://ssc.spitzer.caltech.edu/tools/senspet/} and
do not include confusion with other sources. However, since these
galaxies are not particularly crowded, confusion is not a major source of
uncertainty in the IR flux. The coverage area for all six galaxies is
approximately 5.8\arcmin{} $\times$ 5.8\arcmin{}, centered at the
coordinates listed in Table~\ref{tab:obs}.  Figure~\ref{fig:irfov}
displays the field of view overlaid on the Digitized Sky Survey images
of the six selected galaxies.

GTO and PID 40524 IRAC data were processed with pipelines S14.0.0 and
S16.1.0, respectively, with the exception of the PID 40524 data for
Leo~A, which was processed with pipeline S17.0.4. The Basic Calibrated
Data (BCD) were reduced and mosaicked with the MOPEX\footnote{MOPEX is
available from the {\it Spitzer} Science Center at
http://ssc.spitzer.caltech.edu/postbcd} reduction package
\citep{makovoz05} after applying an array distortion correction.  We
implemented the MOPEX overlap routine to match backgrounds between
overlapping areas of the images and the MOPEX mosaicker for outlier
elimination, image interpolation, and co-addition.  The final mosaics
have pixel sizes of 0.86\arcsec{}~pix$^{-1}$ and are presented in
\citet{jackson06}. The 8~\micron{} mosaics are affected by persistent
images resulting from observations of very bright objects immediately
preceding our program.  These artifacts were removed by subtracting a
median-combined image from each individual frame before proceeding
with MOPEX.

Point-source extraction was performed with the DAOphot~II package
\citep{stetson87}. PSFs were created from a minimum of 10 isolated
stars in each mosaic, and sources 4\,$\sigma$ above the background
were chosen for extraction.  Extended galaxies and outliers that are
broader or narrower than the PSF were rejected based on sharpness and
roundness cut-offs.  The final fluxes are color-corrected
\citep{handbook} using a 5000~K blackbody, which is an appropriate
temperature for a typical Red Giant star. The color correction listed
for a 2000~K star in \citet{handbook} differs by less than 1\%.  A
pixel-phase-dependent correction \citep{reach05} has been applied to
the 3.6~\micron{} photometry.  Photometric errors include standard
DAOphot errors and the IRAC absolute calibration errors
\citep{reach05}. Magnitudes relative to $\alpha$~Lyr (Vega) are
derived using the zero magnitudes quoted in the {\it Spitzer} IRAC
data handbook \citep{handbook}. Table~\ref{tab:photometry} shows a
sample from the full point-source catalog, which is available
electronically and includes IRAC photometry for each of the eight
galaxies in our sample.

\subsection{\emph{Optical Photometry}}
\label{sec:optphot}

We used optical data for each galaxy to aid in source classification,
which is challenging with IRAC data alone (see \S\,\ref{sec:compare}).
Broadband {\it V} and {\it I} images of Sextans~A, Pegasus~dIrr, and
Phoenix were obtained from the Local Group Galaxies
Survey\footnote{LGGS data are publicly available at
http://www.lowell.edu/~massey/lgsurvey/.}
\citep[LGGS;][]{massey06}. The LGGS survey nicely complements IRAC
data because the coverage includes the entire IRAC field of view, and
the angular resolution is similar to the IRAC resolution (median
seeing $\approx$1\arcsec{}).  The published LGGS photometry only
includes sources 10\,$\sigma$ above the background, rendering an
incomplete CMD. In order to reliably identify different regions of the
CMD, it was necessary to recover sources down to 4\,$\sigma$ above the
background. We performed PSF photometry on the LGGS images ourselves
with DAOphot~II. Precise absolute photometry was not required since
the {\it V} and {\it I} data were used only to identify sources in
optical CMDs.  We therefore used ``stacked'' images with relative
photometric errors of $\approx$10\% instead of photometric images
since image processing, including co-addition and astrometric
calibration, had already been performed on the former. To calibrate
the LGGS magnitude offsets between instrumental and true magnitudes,
we compared our photometry to the published LGGS calibrated
10\,$\sigma$ photometry lists. The magnitude offsets and color terms
were computed using least-squares fitting.

{\it V} and {\it I} photometry of DDO~210 and Leo~A were obtained with
the SUBARU telescope by \citet{mcconnachie06} and \citet{vansev04}.
As with the LGGS, the SUBARU coverage includes the entire IRAC field
of view and the angular resolution is similar to the IRAC resolution
(0.8\arcsec{} seeing for both DDO~210 and Leo~A). The calibrated
photometry lists compiled by \citet{mcconnachie06} for DDO~210 and
\citet{vansev04} for Leo~A are sufficiently complete and have relative
photometry errors $\ltsimeq$10\%.

Optical photometry of LGS~3 was obtained with permission from the
Local Cosmology from Isolated Dwarfs Survey
\citep[LCID;][]{gallart08}, which imaged several Local Group Dwarfs
with the ACS camera onboard the {\it Hubble Space Telescope} (HST)
with the F475W and F814W filters.  The total optical coverage for ACS
is $\approx$ 3.3\arcmin{} $\times$ 3.3\arcmin{}
(Fig.~\ref{fig:irfov}). The resolution and PSF of the ACS camera is
vastly superior to those of IRAC, making cross-correlation between the
IR and optical data difficult for LGS~3.  To facilitate comparisons
between optical and IR photometry of AGB candidates in LGS~3, we
inspected the IRAC and ACS images by eye to be sure we found the
correct IR counterparts of the 12 optical point sources detected
in the HST images above the {\it I}-band TRGB.  Due to the inferior
resolution of IRAC, the identified IR counterparts have potentially
been blended with other optical point sources, enhancing the IRAC
flux. However, the 12 optical AGB stars span the entire IRAC
image, leaving no danger of blending between two or more optical AGBs.
The vast majority of the flux from each IR source is therefore
originating from the optically-identified AGB star, since other
optical sources would contribute only very small amounts of flux in
the IR. All optical sub-TRGB sources in LGS~3 remain unmatched to {\it
Spitzer} sources.

\subsection{\emph {Source Contamination}}
\label{sec:foreground}

The main contributors of contamination to the {\it Spitzer} and
optical photometry lists are unresolved background galaxies and
foreground stars. We used the Milky Way stellar population synthesis
model of \citet{robin03} to estimate the degree of contamination in
each IRAC field from the latter. Foreground stellar counts for 1
deg$^2$ fields centered on the Galactic coordinates of each galaxy
were generated to reduce statistical error. The \citet{robin03} model
provides Johnson-Cousins {\it L} band magnitudes, which is similar to
the IRAC 3.6~\micron{} band.  In the IRAC coverage area, we expect to
find 6 foreground stars in the Phoenix and Leo~A mosaics, 10
foreground stars in the Pegasus~dIrr, Sextans~A, and LGS~3 mosaics,
and 32 foreground stars in the DDO~210 mosaic. These stars are
expected to have [3.6] -- [4.5] colors near zero and
$-6\,\ltsimeq\,M_L\,\ltsimeq -20$. The contamination from foreground
stars is therefore only 0.5\% -- 3.5\% of the total number of point
sources detected at 3.6~\micron{}.  This contamination is reflected in
error bars and uncertainties throughout this paper.  Background
galaxies are a more insidious contamination, accounting for anywhere
from 20\% -- 70\% of the sources brighter than the 3.6~\micron{} TRGB
and residing in an IRAC color-magnitude locus similar to that of
obscured AGB stars.  The effect of background contamination on the
analyses presented in this paper is discussed in detail in
\S\,\ref{sec:bkgnd}.

While most of the sources belonging to each galaxy are AGB stars,
there may also be a small population of young stellar objects
(YSOs). YSOs in the Large Magellanic Cloud (LMC) have typical
8~\micron{} magnitudes of $m_{8.0} \approx 10$ mag and have similar
colors to AGB stars, with $2~\ltsimeq~[3.6] - [8.0]~\ltsimeq~4$
\citep{whitney08}. At the distance of the LMC, this puts the vast
majority of YSOs at $M_{3.6} \gtsimeq -6.5$ mag, which is
approximately equal to the TRGB of each of the galaxies in our sample
(see \S\,\ref{sec:TRGB}). Therefore, only the most massive YSOs
($M~\gtsimeq~5~M_\odot$) are included in our AGB statistics. This
contamination is likely very small, given the SFHs of these galaxies.

\section{COMPARISON OF OPTICAL AND IR PHOTOMETRY}
\label{sec:compare}

\subsection{\emph {Optical Color-Magnitude Diagrams}}
\label{sec:optcmd}

\begin{figure}[h!]
\epsscale{.55} \plotone{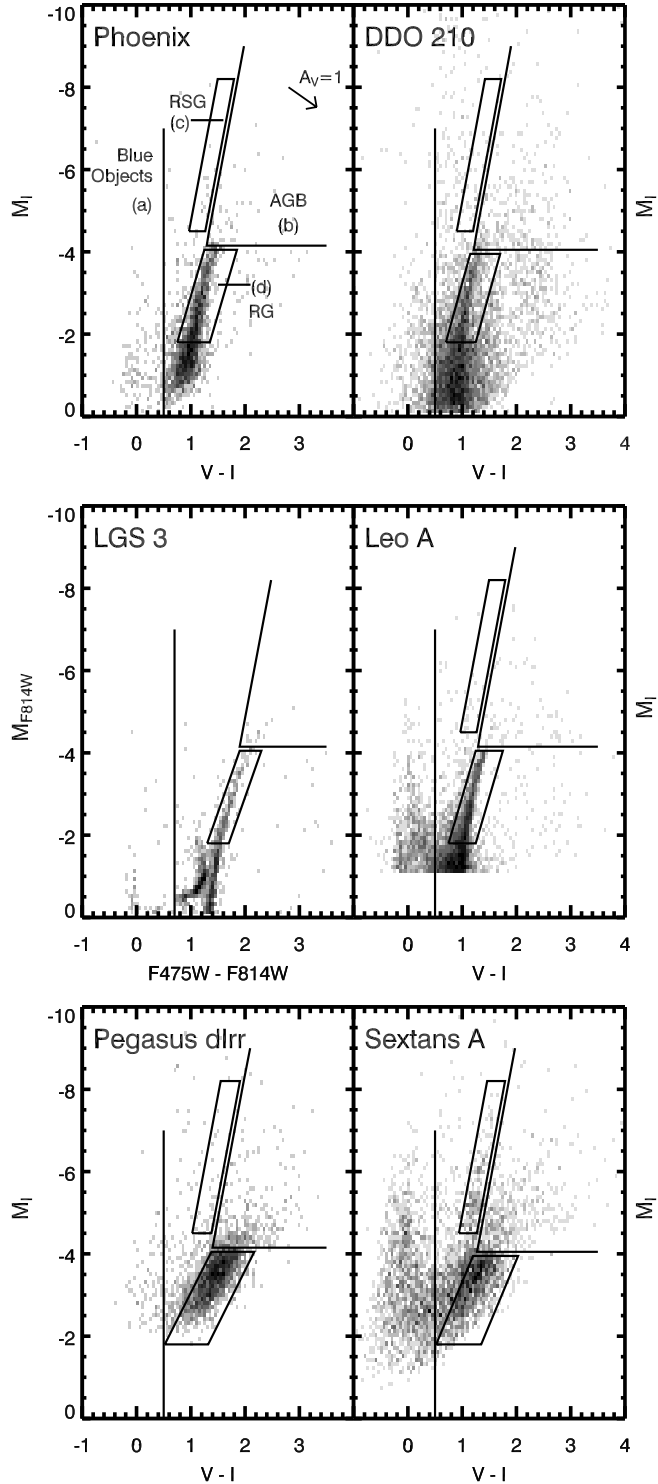} \figcaption{Optical
  color-magnitude diagrams.  See \S\,\ref{sec:optphot} for a
  description of optical photometry.  The color magnitude diagrams are
  represented by Hess diagrams with color bins and magnitude bins of
  0.5~mag. Regions are labeled containing (a) blue objects, (b) AGB
  stars, (c) RSGs, and (d) red giant stars.  These optical
  identifications are used to aid in the identification of different
  source types in the infrared color-magnitude diagrams.  The {\it
  I}-band TRGB is approximately $M_{\rm I} = -4.0$~mag for each
  galaxy. The space between the gaps represent the approximate
  1\,$\sigma$ photometric uncertainties. A vector in the first panel
  shows 1~magnitude of extinction and the associated
  reddening. \label{fig:optcmd}}
\end{figure}

IRAC data sample the Rayleigh-Jeans tail of the Planck function,
providing very little information about stellar effective temperatures
and making it necessary to identify source types in the
optical. Optical CMDs are shown as Hess diagrams in
Figure~\ref{fig:optcmd}. Lines divide the regions where ({\it a}) blue
objects, ({\it b}) AGB stars, ({\it c}) red supergiants (RSGs), and
({\it d}) red giants (RGs) below the TRGB reside. Sources located in
({\it d}) will hereafter be referred to as sub-TRGB stars.  The gap
widths between the regions reflect the 1\,$\sigma$ photometric errors
and reduce misidentification solely due to photometric
uncertainty. Circumstellar and interstellar reddening can also cause
source misidentification that is not corrected for in
Figure~\ref{fig:optcmd}.

The optical CMDs show features that indicate a wide variety of SFHs among these six galaxies.  Sextans~A shows
the strongest evidence for recent and intermediate age star formation
with prominent plumes of blue helium burning stars, RSGs, and AGB
stars that are consistent with an increase in star formation rate
(SFR) by a factor of three 1~Gyr ago and a factor of five 100~Myr ago
\citep{dolphin05,dolphin03b}. Leo~A also shows a prominent branch of
blue helium burning stars, reflecting a twofold increase in the recent
SFR over the lifetime average $\approx$1~Gyr ago \citep{dolphin05}.

Pegasus~dIrr has been forming stars at a moderately high rate over its
lifetime, but the SFR has steadily decreased during the last 1~Gyr.
Blue stars in DDO~210 suggest some star formation 3 -- 6~Gyr ago
\citep{mcconnachie06}, and the SFR has remained steady throughout the
lifetime of the galaxy \citep{dolphin05}. The SFH of Phoenix
\citep{holtzman00, dolphin05, young07} reveals that star formation
ceased $\sim$100~Myr ago.  LGS~3 is similar to Phoenix because it has
been forming stars at a low, but constant level since its initial star
formation event \citep{miller01,dolphin05}, but a lack of blue stars
in LGS~3 suggests very little recent star formation.

\subsection{\emph {IRAC Color-Magnitude Diagrams}}
\label{sec:ircmd}
\begin{figure}[h!]
\epsscale{1} \plotone{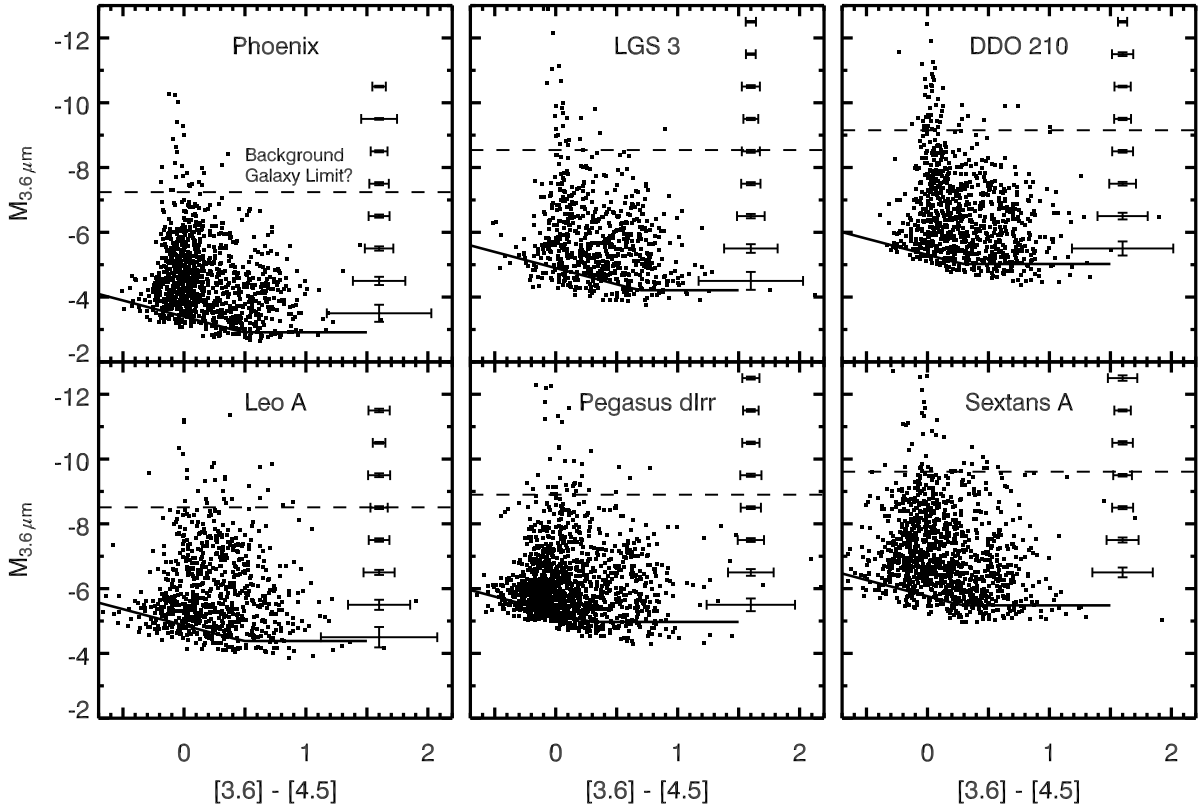} \figcaption{$[3.6]$
  vs. $[3.6] - [4.5]$ color-magnitude diagrams. All point sources
  detected at 3.6 and 4.5~\micron{} are included.  1\,$\sigma$
  photometric errors averaged over 1~mag bins are shown on the right
  of each panel.  The solid lines represent the 50\% completeness
  limits.  The dashed line shows the position of $m_{3.6} = 16$~mag, which
  is the approximate maximum apparent magnitude of point-source
  background galaxies. Red sources above this line are very likely
  obscured AGB stars, since background galaxies brighter than this
  limit are likely extended and subsequently rejected during PSF
  fitting.\label{fig:ircmd}}
\end{figure}

The $M_{3.6}$ versus $[3.6] - [4.5]$ CMDs are shown in
Figure~\ref{fig:ircmd}. These CMDs show all sources detected at 3.6
and 4.5~\micron{} along with the photometric errors, averaged over
one-magnitude bins.  The 50\% completeness limits, averaged over the
entire fields of view, are also shown as a solid black line.  Unlike
WLM, none of the galaxies in this sample are particularly crowded in
the infrared, so the photometric completeness does not decline
significantly toward the centers of the galaxies. Sources that fall
along $[3.6] - [4.5] = 0$ sample the Rayleigh-Jeans tail of the Planck
function at both 3.6 and 4.5~\micron{}. Other sources are reddened by
circumstellar dust emission. The region redward of $[3.6] - [4.5] = 0$
is also populated by background galaxies, although these should mostly
fall below the dashed line (see \S\,\ref{sec:bkgnd}).

The $M_{3.6}$ versus $[3.6] - [4.5]$ CMDs of all eight galaxies
(including WLM and IC~1613 from papers~I and II) are very similar and
differ significantly only in the depth of the photometry. Phoenix has
the fewest bright, red sources ($M_{3.6} < -8, [3.6] - [4.5] > 0.5$),
while WLM, IC~1613, and Sextans~A contain the largest number of
sources in this region. These magnitudes and colors are consistent
with obscured, mass-losing AGB stars (see \S\,\ref{sec:agbs}).

\begin{figure}[h!]
\epsscale{1} \plotone{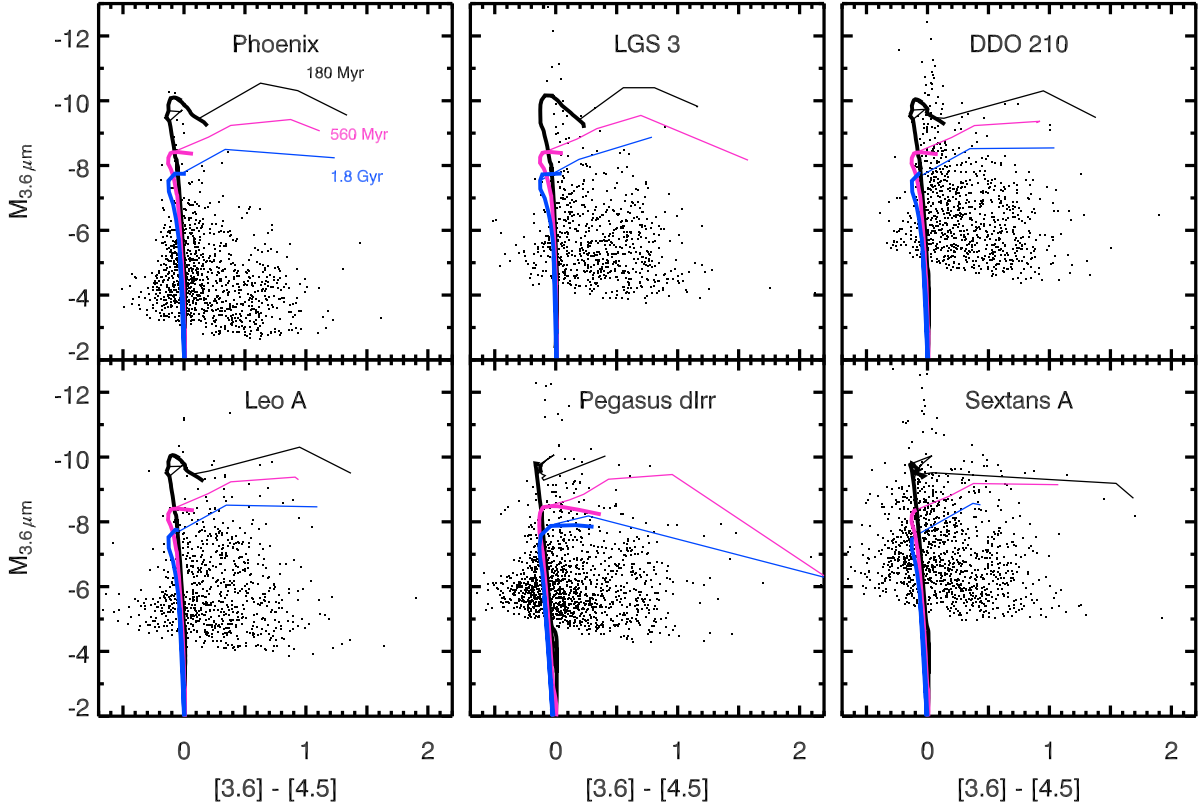}
\figcaption{Color-magnitude diagrams, as in Figure~\ref{fig:ircmd},
with overlaid isochrones from \citet{marigo08}.  Isochrones were
computed at the metallicity of each galaxy for three separate
single-age populations: log($t$) $=$ 8.25 (black), 8.75 (magenta), and
9.25 (blue).  The thick lines represent isochrones for stars with no
circumstellar dust, and the thin lines are isochrones for stars with
dust.  Dust compositions are assumed to be 60\% silicates plus 40\%
aluminum oxides for O-rich stars and 85\% amorphous carbon plus 15\%
SiC for C-rich stars. Mass-losing AGB stars are expected to be among
the brightest and reddest stars in the CMDs.  \label{fig:cmd+iso}}
\end{figure}

Isochrones from \citet{marigo08}, computed for single-aged populations
of log($t$)~$=$~8.25 (black), 8.75 (magenta), and 9.25 (blue) at the
metallicity of each galaxy (Table~\ref{tab:properties}), are overlaid
on the $M_{3.6}$ versus $[3.6] - [4.5]$ CMDs in
Figure~\ref{fig:cmd+iso}.  These isochrones do not perfectly fit the
IR data \citep[e.g., see Figure 3 of][]{whitelock08}, but they do
offer the best representation to date of IR stellar populations due to
the inclusion of dust in the models. The thick lines show the
isochrones that do not include dust, and the thin lines show
isochrones computed with 60\% Silicate $+$ 40\% AlO$_x$ for O-stars,
and 85\% amorphous carbon $+$ 15\% SiC for C-stars. It is clear from these
isochrones that dusty, mass-losing AGB stars are not only red in
color, but are also some of the brightest stars on the CMD.  These
isochrones are not a good match to faint, red stars ($M_{3.6} \gtrsim
-7~\rm{mag}, [3.6] - [4.5] \gtrsim 0.5$).  This discrepancy may be due
to a combination of photometric uncertainty in this region of the CMD
and contamination from other source types, particularly background
galaxies.

\begin{figure}[h!]
\epsscale{1} \plotone{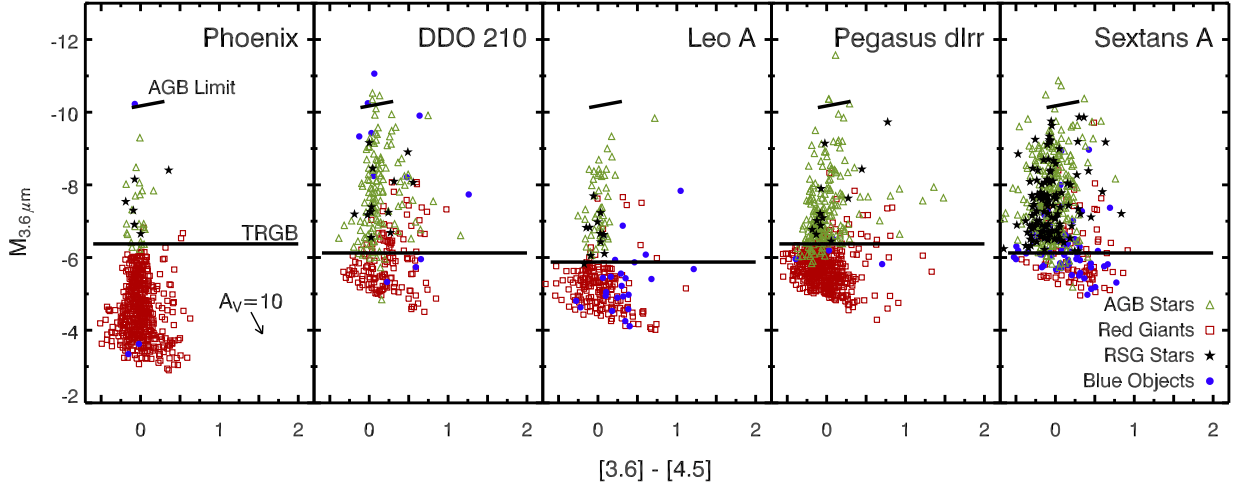} \figcaption{Optically identified sources
  in the $[3.6]$ vs. $[3.6] - [4.5]$ color-magnitude diagrams.  Red
  squares are red giants, black stars are RSGs, blue circles are blue
  objects, and green triangles are optical AGB stars, as identified in
  Figure~\ref{fig:optcmd}.  The first panel displays a vector showing
  10 visual magnitudes of extinction and the associated
  reddening. Note that there is virtually no extinction at
  3.6~\micron{}. Each panel shows the AGB limit and the location of
  the TRGB. Note that red stars above the TRGB, evident if
  Figures~\ref{fig:ircmd} and \ref{fig:cmd+iso}, are largely
  undetected in the optical. \label{fig:colorcmd}}
\end{figure}

Figure~\ref{fig:colorcmd} shows point-sources residing in regions
({\it a}), ({\it b}), ({\it c}), or ({\it d}) of the optical CMDs
(Fig.~\ref{fig:optcmd}) of Phoenix, DDO~210, Leo~A, Pegasus~dIrr, and
Sextans~A; red squares are red giants, green triangles are optical AGB
stars, black stars are RSGs, and blue circles are blue objects. LGS~3
is excluded from Figure~\ref{fig:colorcmd} due to difficulty in
matching the {\it Spitzer} sources to the ACS sources as discussed in
\S\,\ref{sec:optphot}. Since there is virtually no extinction at
3.6~\micron{}, we show a vector in the first panel of
Figure~\ref{fig:colorcmd} illustrating the displacement corresponding
to 10~magnitudes of visual extinction and the associated reddening
\citep{rieke85,indebetouw05}. This vector does not take into account
the effect of dust emission, which can be considerable in the
infrared.  The AGB limit was determined by scaling AGB stars with
$T_{\rm eff} = 2650$ and 3600~K and negligible mass loss
\citep{groenewegen06} to bolometric magnitudes of $M_{\rm bol} =
-7.1$~mag. The lines shown in Figure~\ref{fig:colorcmd} connect the
limits of these two hypothetical stars. AGB stars that are undergoing
heavy mass loss can attain 3.6~\micron{} fluxes brighter than this
limit due to thermal emission from the expelled material.  Such stars
would also have red $[3.6] - [4.5]$ colors, as is the case for one
star in Leo~A (Fig.~\ref{fig:ircmd}).  All other stars above the AGB
limit in our galaxy sample are likely RSGs. Note that few of the stars
following the isochrones in Figure~\ref{fig:cmd+iso} are detected in
the optical, especially in Leo~A. Moreover, of those stars that are
detected in $V$ and $I$, many are not {\it optically} identified as
AGB stars and are either contaminants or obscured AGB stars.

\subsubsection{The Tip of the Red Giant Branch}
\label{sec:TRGB}

\begin{figure}[h!]
\epsscale{.75} \plotone{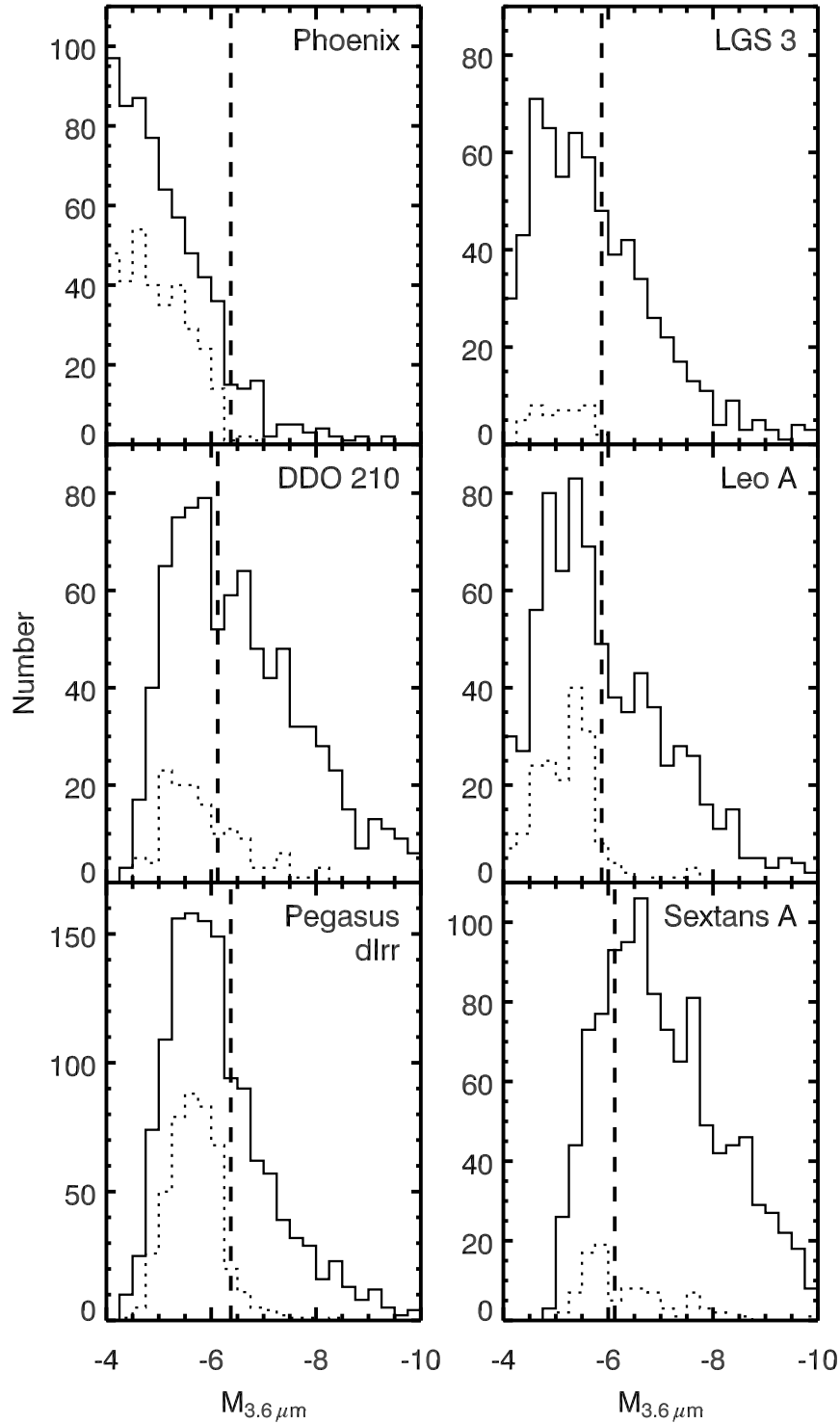}
  \figcaption{3.6~\micron{} luminosity functions. The solid line
  represents all sources detected at 3.6 {\it and} 4.5~\micron{}. The
  dotted line shows optically classified sub-TRGB stars. The adopted
  TRGB values are marked with a dashed line. In the case of Sextans~A,
  where the IRAC luminosity function drops off at a magnitude brighter
  than the TRGB, the optical sub-TRGB luminosity function provides a
  first-order estimate of the 3.6~\micron{} TRGB. \label{fig:lfunc}}
\end{figure}

Figure~\ref{fig:colorcmd} also shows the locations of the
3.6~\micron{} TRGB.  The TRGB was identified by plotting the
luminosity functions of stars detected at both 3.6 and 4.5~\micron{}
(Fig.~\ref{fig:lfunc}) and determining the magnitude where the source
count decreases significantly. We adopt TRGB values of $M_{3.6} =
-6.38 \pm 0.25$, $M_{3.6} = -5.88 \pm 0.25$, $M_{3.6} = -6.13 \pm
0.25$, $M_{3.6} = -5.88 \pm 0.25$, and $M_{3.6} = -6.38 \pm 0.25$~mag
for Phoenix, LGS~3, DDO~210, Leo~A, and Pegasus~dIrr,
respectively. The dotted lines in Figure~\ref{fig:lfunc} represent the
IR luminosity function of sub-TRGB stars identified in the optical.
In each case, the number counts of the sub-TRGB stars support what we
have determined is the location of the TRGB.  Sub-TRGB stars that are
brighter than the 3.6~\micron{} TRGB have very red colors that are
consistent with heavily obscured, mass-losing AGB stars and might be
reddened enough in the optical to be misidentified as sub-TRGB. 

The IRAC photometry of Sextans~A is not deep enough to reliably
determine the 3.6~\micron{} TRGB, but the luminosity function of the
sub-TRGB stars seems to suggest a TRGB of approximately $M_{3.6} =
-6.13$~mag.  In papers~I and II, we found 3.6~\micron{} TRGBs of
$M_{3.6} = -6.6 \pm 0.2$~mag for WLM and $M_{3.6} = -6.2 \pm 0.2$~mag
for IC~1613. Most of our values are a few tenths of a magnitude
fainter than the TRGB found in the LMC \citep[$M_{3.6} =
-6.65$~mag;][]{blum06}, but this discrepancy is not necessarily
unexpected since different SFHs will yield different TRGBs due to
varying ages and metallicities of RGB stars. The 3.6~\micron{} TRGBs
in old star clusters in the Magellanic Clouds is $M_{L'} \approx
-6$~mag \citep{vanloon05}, which is more consistent with the TRGB
values measured here.

\subsubsection{8~\micron{} Color-Magnitude Diagrams}
\label{sec:8cmd}

\begin{figure}[h!]
\epsscale{.7} \plotone{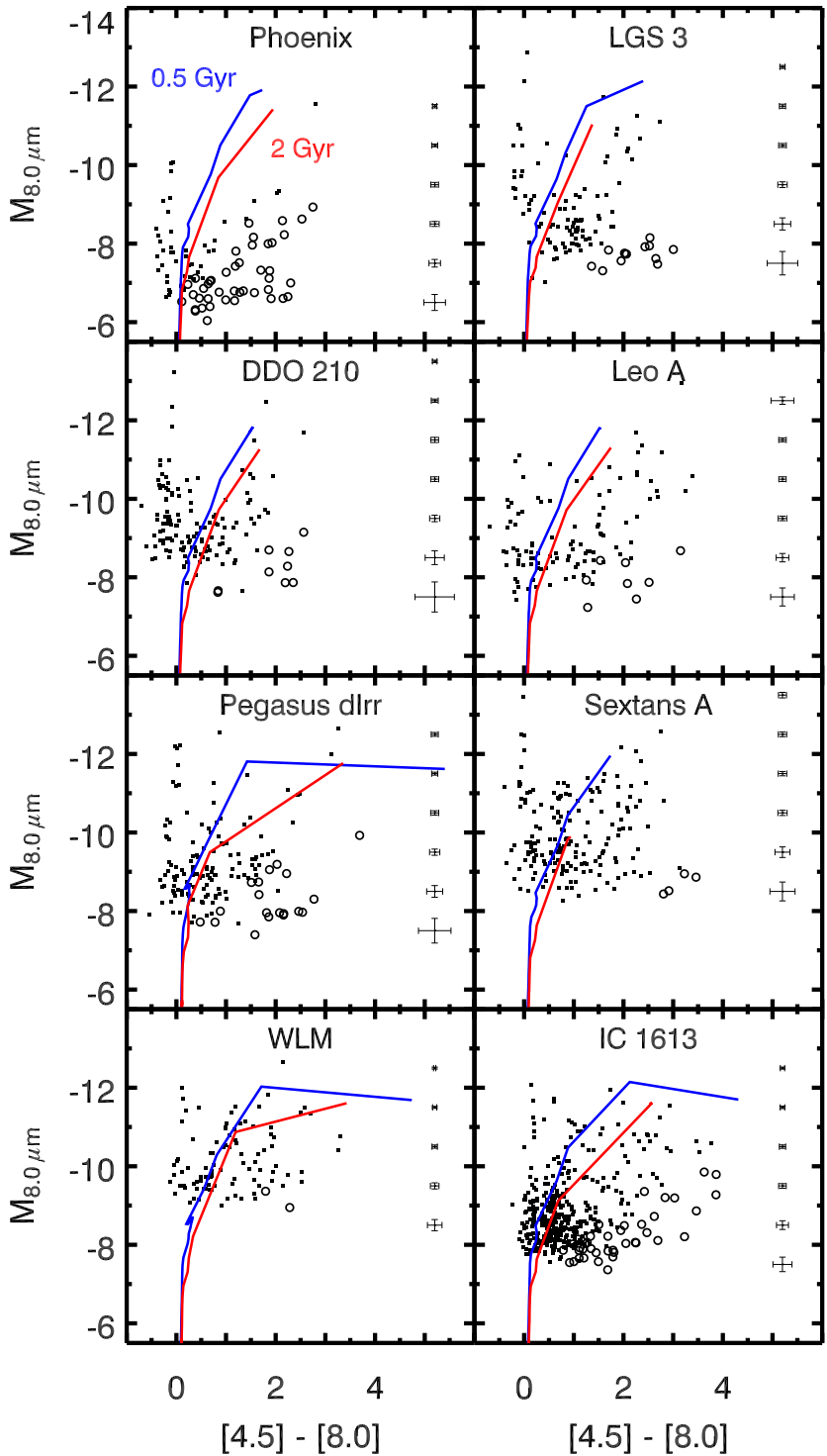} \figcaption{$[8.0]$
  vs. $[4.5] - [8.0]$ color-magnitude diagrams.  Photometric
  1\,$\sigma$ errors, averaged over 1~magnitude bins, are shown on the
  right side of each panel.  Open circles mark sources that are below
  the 3.6~\micron{} TRGB.  In each galaxy, a plume of bright, red
  stars generally falls just redward of the isochrones for mass-losing
  AGB stars \citep{marigo08}. The isochrones were computed using the
  metallicities listed in Table~\ref{tab:properties}, and a single-age
  population of 2~Gyr (red line) and 0.5~Gyr (blue line). Assumed dust
  compositions are identical to those in
  Figure~\ref{fig:cmd+iso}. \label{fig:cmd14}}
\end{figure}

The $M_{8.0}$ versus $[4.5] - [8.0]$ CMD is presented in
Figure~\ref{fig:cmd14}. In these CMDs, sources below the 3.6~\micron{}
TRGB are plotted as open circles.  As with Figure~\ref{fig:ircmd},
this CMD is similar for all eight galaxies.  The main features are a
narrow plume centered at $[4.5] - [8.0] = 0$ and a bright, red plume
with $M_{8.0} < -8.0$~mag and $[4.5] - [8.0] > 1.0$. The red plume is
strongly populated in all eight galaxies, with the exception of
Phoenix, which has fewer than 10 stars in this region that are
brighter than the TRGB. Based on the colors of background galaxies in
$\omega$\,Cen \citep{boyer08}, the reddest regions of the CMDs in
Figure~\ref{fig:cmd14} are contaminated by background galaxies ($[4.5]
- [8.0] > 2$), although the most heavily obscured AGB stars can also
reach colors as red as $[4.5] - [8.0] \approx 5$. 

Isochrones for two different ages are overlaid in
Figure~\ref{fig:cmd14} \citep[$\approx$0.5 and 2~Gyr;][]{marigo08} and
demonstrate that obscured AGB stars are bright at 8~\micron{}.
Because the isochrones appear to be systematically bluer than the
data, we do not attempt a fit, but it is clear that the locus of the
bright, red plume of sources in each galaxy is consistent with
obscured AGB stars, while fainter red sources may be other
source-types such as background galaxies or YSOs.

\subsection{\emph {Detection Statistics}}
\label{sec:detstats}

The detection statistics for the regions shared by optical and IRAC
photometry are presented in Table~\ref{tab:detstats} (note that the
LGS~3 optical coverage is smaller than the IR coverage due to the
small field of view of the ACS camera). We consider a
source to be detected in {\it V} and {\it I} only if it is brighter
than $M_{\rm I} = -2.5$~mag, which is the approximate limiting magnitude
of the shallowest optical data. Sextans~A has the largest 8~\micron{}
population, followed by Pegasus~dIrr. The number of detections in the
IRAC data of IC~1613 and WLM presented in papers~I and II are much
higher than all six of the galaxies presented here, but that is due to
an increased areal coverage (more IRAC fields) rather than to an
intrinsic property of the galaxies themselves.

In each galaxy, we find between 27 -- 67 sources that are detected in
all four IRAC bands, but not in {\it V} and {\it I}.  The majority of
these sources lie between the TRGB and the AGB limit and may be either
optically obscured AGB stars or background galaxies.  We are unable to
determine the corresponding numbers for LGS~3 due to the difficulty of
matching IRAC data to ACS data (see \S\,\ref{sec:optphot}). However,
we have inspected the bright optical sources by eye and note that
there are a mere 12 stars brighter than the {\it I}-band TRGB in LGS~3
that are also detected in IRAC. For comparison, in the IRAC images,
there are 102 sources located between the 3.6~\micron{} TRGB and the
AGB limit in LGS~3 in the area covered by the ACS data.

A small number of the IRAC detections with no optical counterparts are
above the AGB limit and have colors near zero. These bright sources
are probably foreground stars or RSG stars, as any AGB star above the
AGB limit would have a very red color.

There is strong evidence for confusion by background galaxies in our
sample (see \S\,\ref{sec:bkgnd}), especially below the 3.6~\micron{}
TRGB. Phoenix, the nearest of the eight galaxies in our sample,
hosts the largest number of sources that are below the 3.6~\micron{}
TRGB and not detected in the optical (25 sources total;
Table~\ref{tab:detstats}).  The region occupied by background galaxies
in Figure~\ref{fig:ircmd} is at the faintest absolute magnitude for
Phoenix, pushing many of the background galaxies below the
3.6~\micron{} TRGB.  As a result, Phoenix has the smallest number of
background galaxies {\it above} the TRGB at 3.6~\micron{}.

\subsection{\emph {Background Galaxy Contamination}}
\label{sec:bkgnd}

Unresolved background galaxies tend to have [3.6] -- [4.5] colors very
similar to optically obscured AGB stars.  Based on inspection of IRAC
CMDs alone, background galaxies are easily identified in observations
of nearby stellar populations such as the Magellanic Clouds
\citep{blum06, bolatto07} and $\omega$\,Centauri \citep{boyer08}
because the AGB stars in these nearby systems have far brighter
apparent magnitudes than the background objects. Unfortunately, the
distances to our target dIrr galaxies are just large enough so that
dusty AGB stars and unresolved background galaxies occupy the same
color-magnitude space and are impossible to disentangle, potentially
causing an overestimate of the dust production and integrated
mass-loss rate in each galaxy.  Because WLM and IC~1613 have the
highest stellar surface densities in the sample, the effects of
unresolved background galaxies were more subtle and not dealt with in
papers~I and II. The lower stellar surface densities of the galaxies
in the new sample makes the effects of unresolved background galaxies
more obvious, and thus we can address these effects in this study.

In the {\it Spitzer} observations of $\omega$\,Centauri
\citep{boyer08}, the unresolved background galaxies remained confined
to 3.6~\micron{} apparent magnitudes fainter than $\approx$16 mag and
colors $-0.3 \gtrsim [3.6] - [4.5] \lesssim 1.0$.  This limit suggests
that we may be confident that any red point-sources brighter than
$m_{3.6} = 16$ mag are mass-losing AGB stars.  Below this limit, we
cannot distinguish between mass-losing AGB stars, background galaxies,
and other red IR sources (e.g., YSOs) with IRAC data
alone. It is clear in Figure~\ref{fig:ircmd} that the pack of red
sources fainter than $m_{3.6} = 16$ mag has a similar shape in each
galaxy and slides up and down with respect to the rest of the IR
population, depending on the distance to the galaxy. Phoenix, which is
the nearest galaxy in our sample, is the least affected by
contamination since the background galaxy limit is only one magnitude
brighter than the TRGB.  In Sextans~A, on the other hand, the region
occupied by galaxies reaches up to 3.5 mag brighter than the TRGB.

In fact, if we assume that the background galaxy populations in the
images of Phoenix, LGS~3, DDO~210, Leo~A, Pegasus~dIrr, and Sextans~A
are the same (the IRAC images for these six galaxies have the same
areal coverage), we should find approximately the same number of
sources fainter than $m_{3.6} = 16$~mag and redder than $[3.6] - [4.5]
\approx 0$. We created Hess diagrams with 0.1~mag color bins and
0.5~mag magnitude bins for Figure~\ref{fig:ircmd} and subtracted the
Phoenix Hess diagram from the others, since the Phoenix background
population is the least mixed with real AGB stars. We find that, in
the region of contamination, the subtraction leaves only $\pm10$
sources in each bin, suggesting that the majority of sources in this
region of the CMD are in fact background galaxies.

In order to obtain a first-order estimate of the degree of
contamination above the TRGB in each of our targets, we inspected the
IRAC CMDs produced with photometry from the {\it Spitzer}-Cosmic
Evolution (S-COSMOS) database \citep{sanders07}, which is a {\it
Spitzer} Legacy survey covering a 2 deg$^2$ dark portion of the
sky. S-COSMOS is complete to more than one magnitude below the
3.6~\micron{} TRGB of our most distant galaxy, allowing a direct
comparison of the background galaxies in S-COSMOS to the number of and
distribution of sources above the TRGB in each galaxy in our
sample. We applied the distance modulus of each galaxy to the S-COSMOS
CMD and counted sources with magnitudes brighter than the TRGB and
$[3.6] - [4.5]$ colors redder than 0.2.  After correcting for the
field size, we find that background galaxies may account for anywhere
from 21\% to 50\% of the sources brighter than the 3.6~\micron{} TRGB
(Table~\ref{tab:detstats}).

For targets that lie in the direction towards a large galaxy cluster,
the galaxy counts can be significantly larger.  LGS~3 and IC~1613 are
each located within 30\arcmin{} from the center of one galaxy cluster
and Leo~A and WLM are each located within 30\arcmin{} from the centers
of two such clusters \citep[see Table~\ref{tab:galclust} for a list of
galaxy clusters within 1 degree of our
targets;][]{abell95,fernandez96}.  None of these clusters are
particularly large or heavily populated, but their presence could
potentially double the number of background galaxies above the average
population.  Moreover, LGS~3 is projected onto the fringes of the
Pisces-Pegasus Supercluster of galaxies, which may cause an even
larger increase in the background galaxy counts in the LGS~3
images. 

In order to measure the background galaxy contamination directly from
the data, a large field of view is necessary.  A search of the {\it
Spitzer} archive revealed IRAC off-fields for Phoenix, DDO~210, Leo~A,
and Pegasus~dIrr as part of the Large Volume Legacy Survey
\citep{kennicutt07}, but the depth of these observations is less than
20\% the depth of our observations, preventing reliable photometry of
sources down to the TRGB.

\begin{figure}[h!]
\epsscale{.48} \plotone{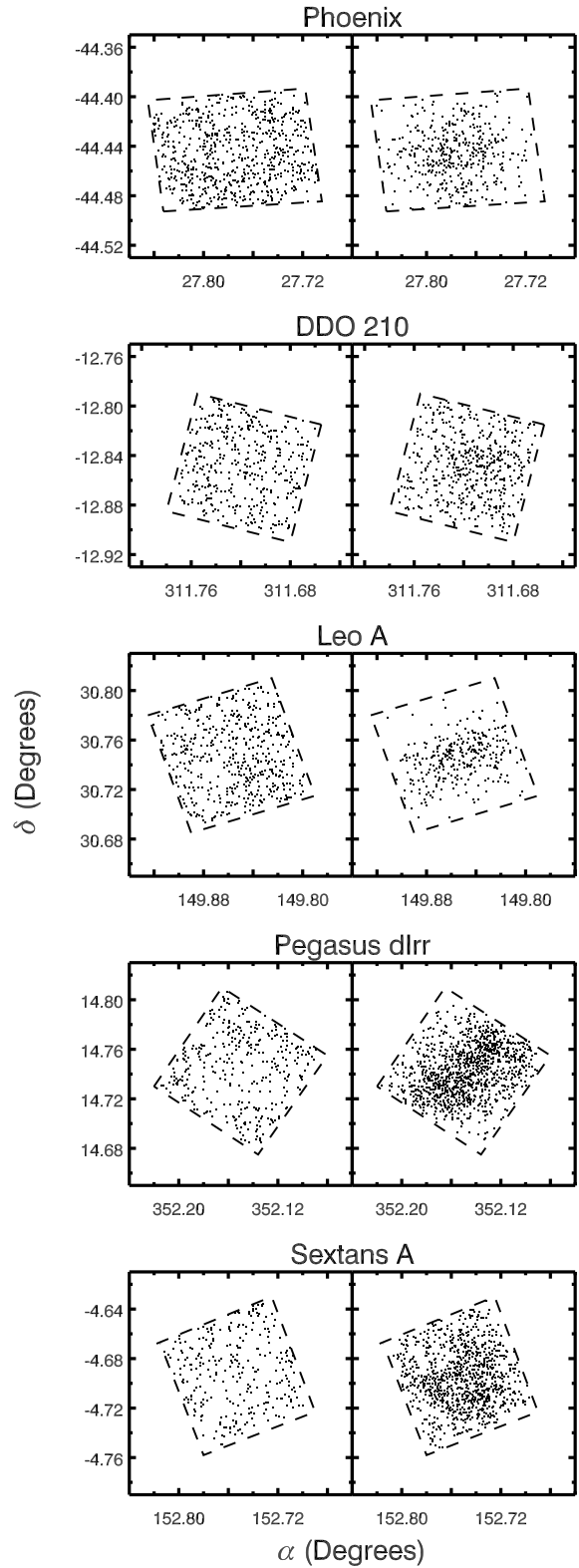} \figcaption{Spatial
distributions of {\it all} IR sources without ({\it left}) and with
({\it right}) optical counterparts, where a source is considered
undetected in the optical if $M_{\rm I} > -2.5$~mag. Sources without
optical counterparts are likely either dust-enshrouded AGB stars or
background galaxies (and possibly YSOs).  The sources
in the left panels do not cluster towards the centers of the galaxies,
suggesting at least some of them are not truly AGB
stars.\label{fig:oidista}}
\end{figure}

\begin{figure}[h!]
\epsscale{1} \plotone{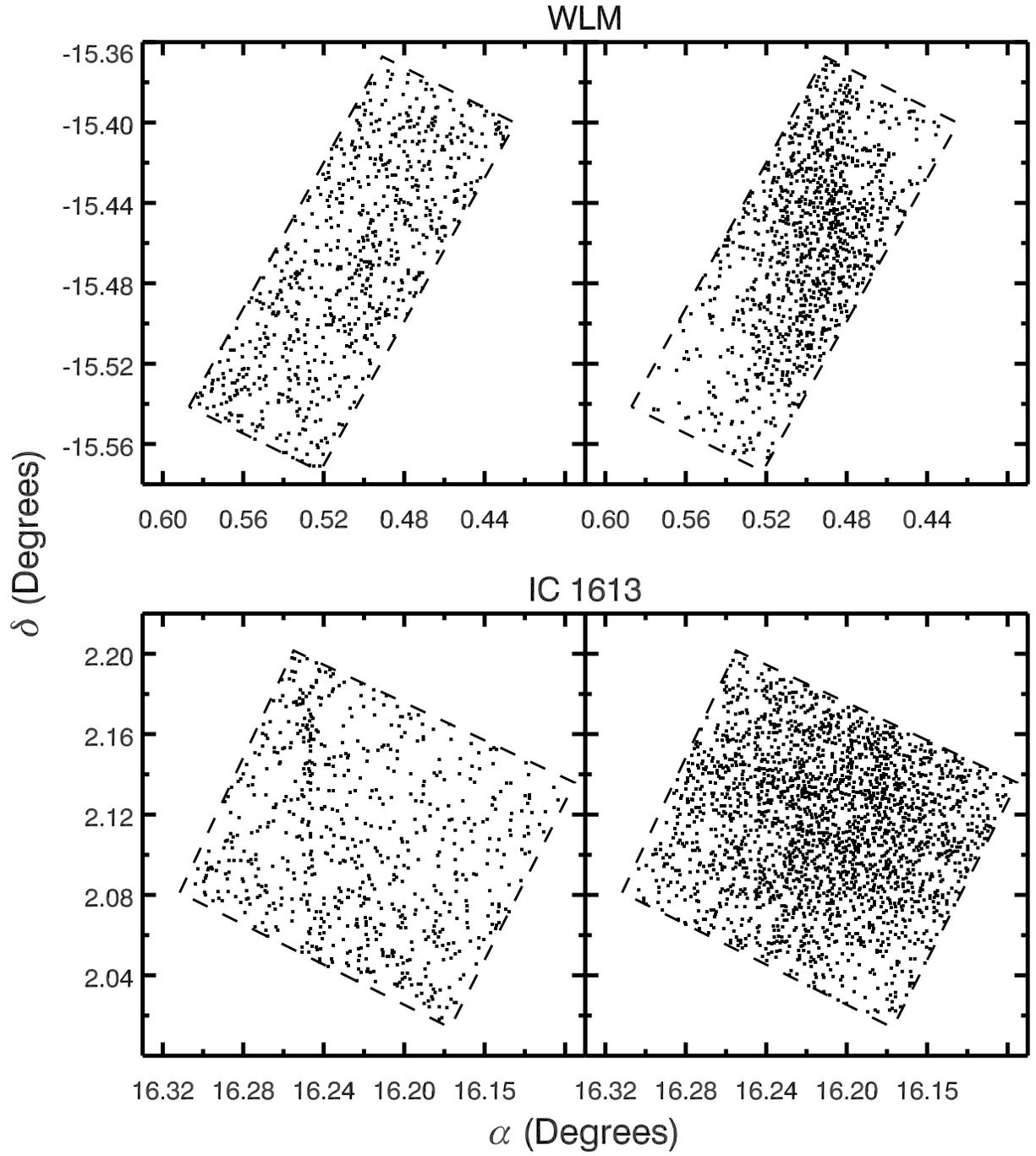} \figcaption{Same as
Figure~\ref{fig:oidista} for WLM and IC~1613.\label{fig:oidistb}}
\end{figure}

Figures~\ref{fig:oidista} and \ref{fig:oidistb} display the spatial
distributions of IR sources with ({\it right}) and without ({\it
left}) optical counterparts. While not completely avoiding the central
parts of each galaxy, IR sources without optical counterparts are more
numerous on the outskirts, where unresolved background galaxies are
expected to begin to dominate the source counts.  Despite the small
fields of view of the galaxies in our sample, it is clear from
Figures~\ref{fig:oidista} and \ref{fig:oidistb} that background
galaxies do, sometimes significantly, contaminate the IR source
count. We attempt to measure the contribution from background galaxies
in each of our targets using the radial density profiles, a task that
would be more robust given a larger field of view but that nonetheless
provides interesting results with the available data.

\begin{figure}[h!]
\epsscale{.8} \plotone{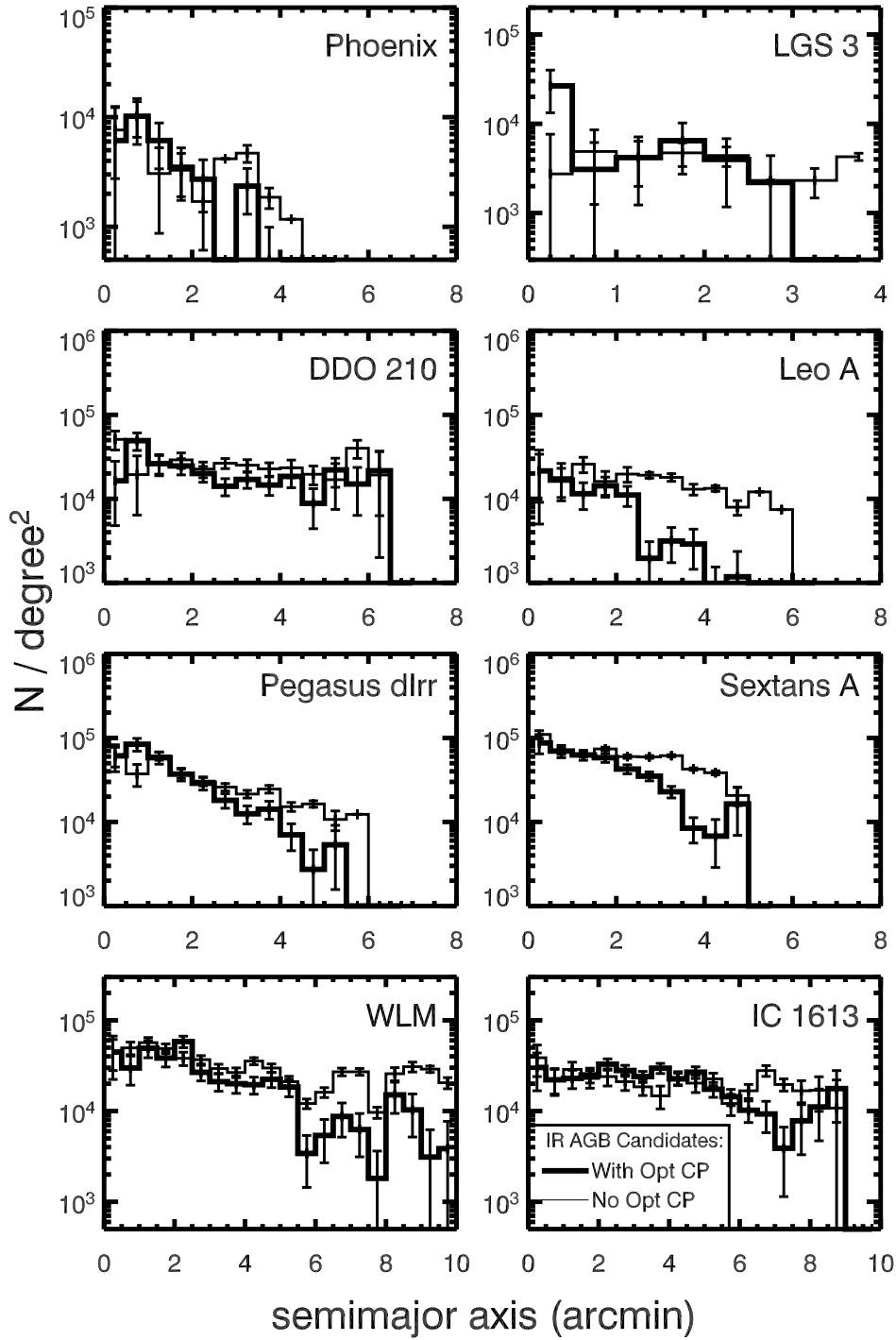} \figcaption{Radial density
profiles for IR AGB candidates with optical counterparts (``Opt CPs'')
brighter than (thick line) and fainter than (thin line) the {\it
I}-band TRGB, with the latter normalized to the former. Profiles were
determined using ellipsoidal annuli with semi-major axis bins of
0.5\arcmin{} centered on each galaxy. If these two populations are
obscured and non-obscured AGB stars, the two profiles should be
identical. However, it is clear that a flat distribution of background
galaxies is contaminating most of our targets. We have fit the
declining profile plus a flat distribution to each profile for sources
without optical counterparts to measure the background contamination,
but the uncertainties in the fits are large and would be much improved
by observing larger fields of view. \label{fig:profiles}}
\end{figure}

Radial density profiles of sources with and without optical
counterparts and brighter than the TRGB are shown in
Figure~\ref{fig:profiles}, with the latter population normalized to
the former.  The profiles were determined by placing ellipsoidal
annuli on each galaxy at semi-major axis intervals of 0.5\arcmin. We
have assumed that the radial profiles should have similar shapes, and
that any deviation is due to a flat contribution from background
galaxies.  Leo~A shows the largest difference between the profiles, as
might be expected considering the number of background galaxy clusters
in that direction (Table~\ref{tab:galclust}). DDO~210 shows the
flattest overall distribution, which suggests that, for this galaxy,
there may be significant contamination in the optical as well as the
infrared.

Using a least squares fitting routine, we have fit a combination of a
declining profile (the profile for AGB candidates with optical
counterparts) plus a flat distribution (background galaxies) to the
profile of AGB candidates without optical counterparts. The resulting
fits agree with the predictions from S-COSMOS to within 25\% for four
galaxies (Phoenix, Leo~A, Pegasus~dIrr, and WLM), 40\% for three
(LGS~3, DDO~210, and Sextans~A), and 78\% for IC~1613. The resulting
total background contamination for the galaxies in our sample ranges
from 25\% to 67\% of the sources above the TRGB (compared to 21\% --
50\% from S-COSMOS).  For DDO~210, the fit has increased the number of
galaxies predicted by S-COSMOS by 37\%, resulting in a larger number
of background galaxies than there are optical AGB stars.  This
suggests that either there are no obscured AGB stars in DDO~210, or
that DDO~210 is significantly contaminated in the optical, as well as
in the infrared.

\subsection{\emph {Stellar Spatial Distributions}}
\label{sec:distributions}
\begin{figure}[h!]
\epsscale{1} \plotone{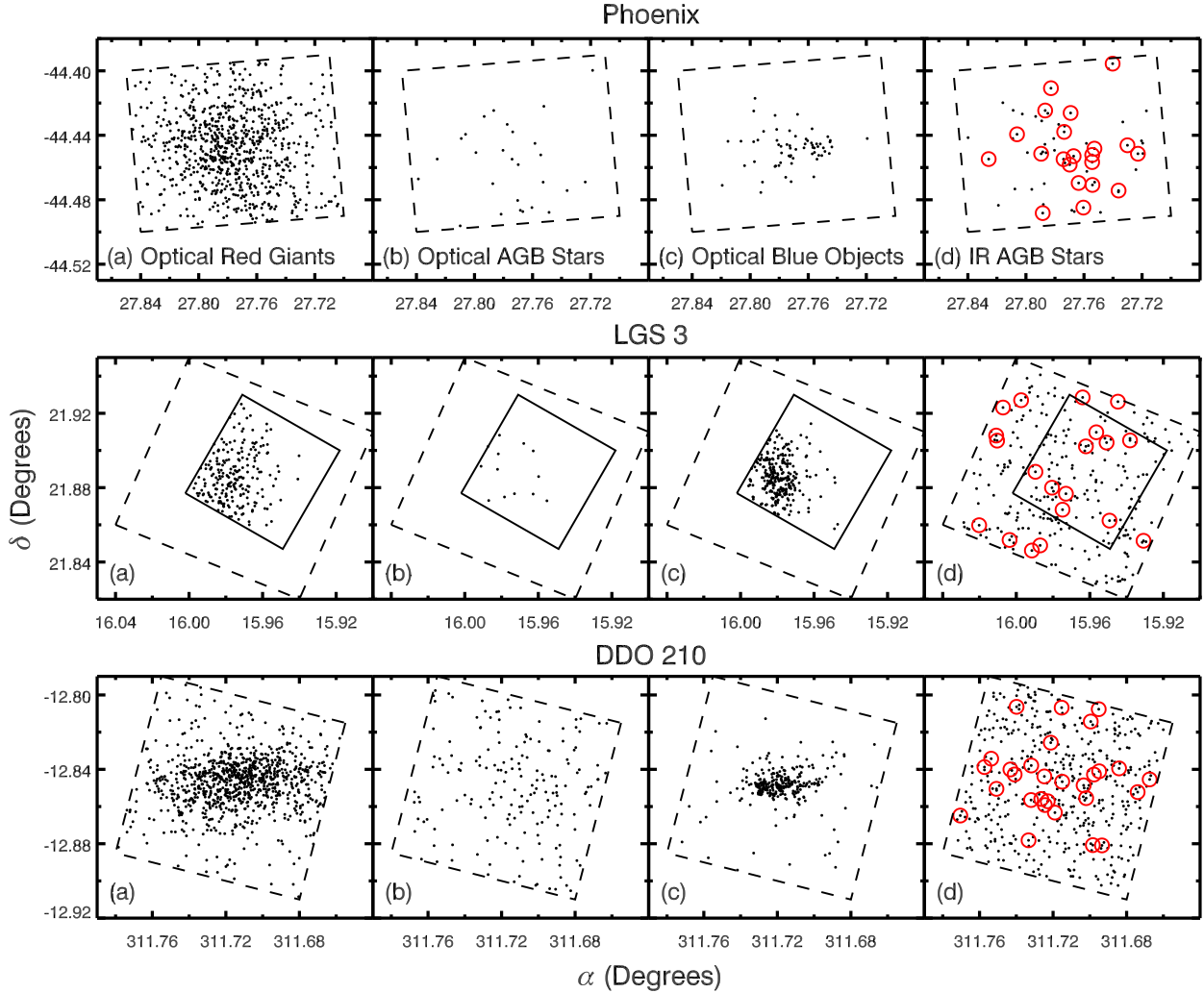} \figcaption{Stellar spatial
  distributions of (a) optical red giants, (b) optical AGB stars, (c)
  optical blue objects, and (d) IR-identified AGB candidates in
  Phoenix, LGS~3, and DDO~210.  The dashed lines outline the {\it
  Spitzer} coverage. The solid box in each LGS~3 panel shows the area
  covered by optical data. In each panel (d), AGB candidates brighter
  than $m_{3.6} = 16$~mag are plotted with open red circles in order to
  show the distribution of AGB stars with little to no contamination
  from background galaxies.\label{fig:dist1}}
\end{figure}

\begin{figure}[h!]
\epsscale{1} \plotone{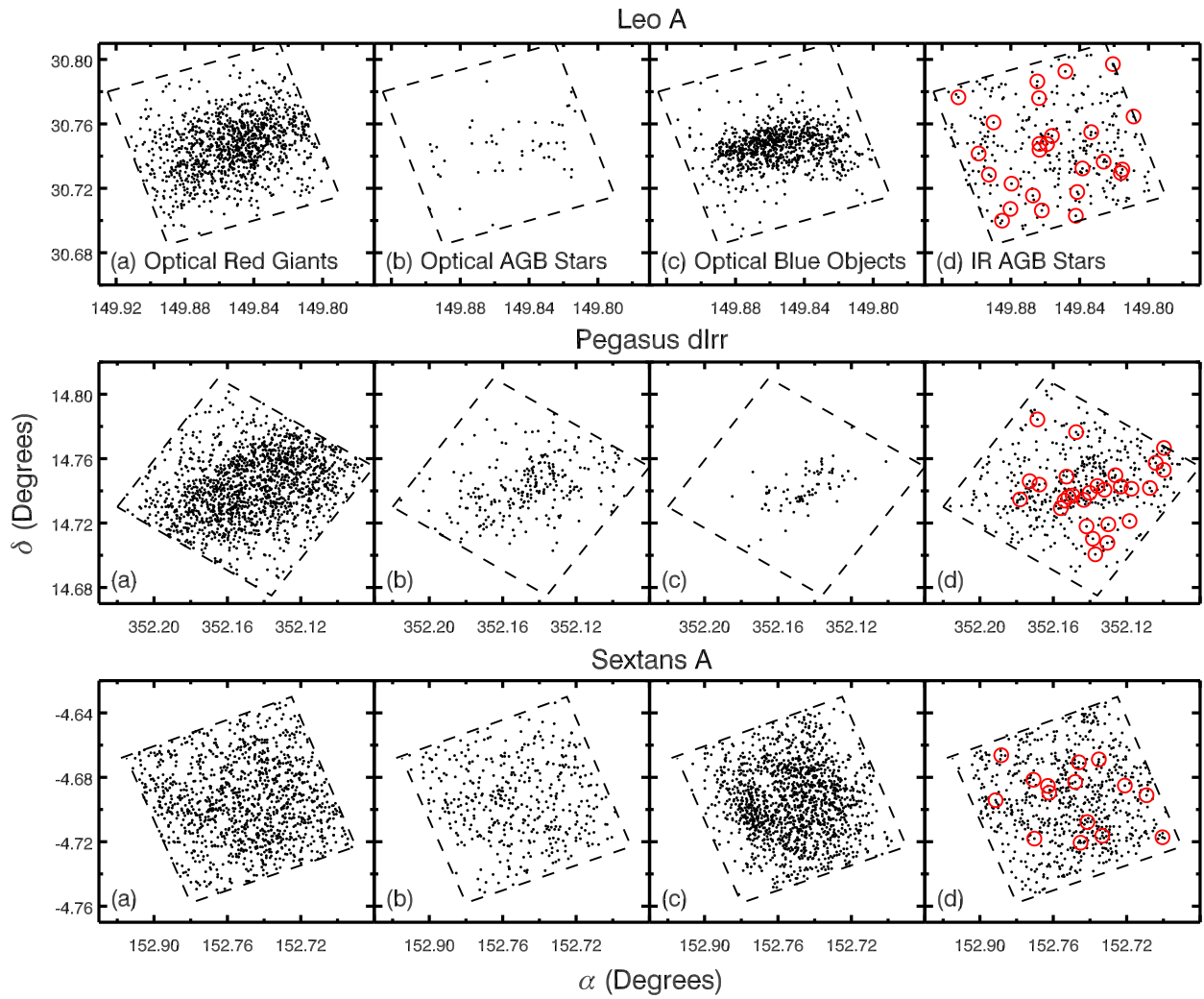} \figcaption{Stellar spatial
  distributions for Leo~A, Pegasus~dIrr, and Sextans~A. The panels are the
  same as in Figure~\ref{fig:dist1}.  \label{fig:dist2}}
\end{figure}

The spatial distributions of different stellar types are displayed in
Figures~\ref{fig:dist1} and \ref{fig:dist2}.  From left to right, the
panels show the optically-identified red giant stars, AGB stars, and
blue objects, followed by the IR-identified AGB star candidates, which
we define as any object between the TRGB and the AGB limit that is not
identified as an RSG or a blue object in the optical
(\S\,\ref{sec:agbs}). The analogous figure in paper~I for WLM shows the
effects of crowding, illustrated by a deficit of fainter red giants in
the regions densely populated with blue stars (i.e., the regions
containing the youngest stellar populations). Similar crowding is not
seen in IC~1613 (Paper~II) nor in any of the six other galaxies
presented here due to a combination of a lower density of stars and
smaller inclination angle \citep[each galaxy, aside from WLM, has $i <
55\degr$;][and references therein]{mateo98}. It is clear from
Figures~\ref{fig:dist1} and \ref{fig:dist2} that the IRAC data do not
cover the entire areal spans of all six galaxies since the stellar
densities are still decreasing at the edges of the coverage. 

The few IR AGB candidates detected in Phoenix span a similar area to
their optically-identified counterparts. Red giants and blue objects
in Pegasus~dIrr and DDO~210 trace a line from east to west across the
fields of view, but the IR AGB candidates are more smoothly
distributed. The brightest IR AGB candidates (red circles) in both
galaxies tend to cluster more towards the center. Sextans~A shows a
mildly clumpy structure with the IR AGB candidates loosely tracing the
young blue objects and RGB stars. Optical studies of the spatial
distributions of the stellar populations are available from HST
studies of Pegasus~dIrr \citep{gallagher98} and Sextans~A
\citep{dohm-palmer97, dohm-palmer02}.  In Pegasus~dIrr, the youngest
stars are strongly clumped, while the AGB stars tend to follow the
distribution of the RGB stars.  In Sextans~A, again, the youngest
stars are strongly clumped, but the RGB stars are both centrally
concentrated and form a smooth halo-like population.  Neither the
optically identified nor IR AGB stars in Figure~\ref{fig:dist2} tend
to be concentrated in the bar feature. This is consistent with the
expectation that the AGB stars show spatial distributions intermediate
between those of the youngest and oldest stellar populations.

For LGS~3, it is difficult to draw connections between the
distributions of the different stellar types due to the small optical
coverage. However, in LGS~3, the distribution of IR AGB candidates is
generally smooth across the entire IRAC field of view, with no
increase in the number of sources in the regions of highest stellar
density. This distribution supports the claim made in
\S\,\ref{sec:bkgnd} that LGS~3 is strongly contaminated by background
galaxies. However, even the brightest IR AGB candidates ($m_{3.6} <
16$~mag), marked with red circles in Figure~\ref{fig:dist1}, span the
entire LGS~3 coverage, suggesting that AGB stars have a broader
distribution than the Red Giant Branch (RGB) stars. Leo~A, which is
also potentially strongly contaminated by background galaxy clusters,
has a slightly higher density of IR AGB candidates on the west edge of
the field, which is offset from the centers of the blue object and
optical red giant distributions.  None of the known background galaxy
clusters (Table~\ref{tab:galclust}) are located near this
overdensity. The brightest AGB candidates in Leo~A are more broadly
distributed (red circles). In both Leo~A and LGS~3, the broad
distributions of AGB stars are not likely to be a reflection of the
true distributions, but rather the result of low number statistics in
these lightly populated galaxies. However, we note a recent study of
variable stars in the Fornax dwarf galaxy \citep{whitelock08} that
suggests that AGB stars in that galaxy are more broadly distributed
than the bulk population.

\subsection{\emph {Luminosity Contribution from Super-TRGB Stars}}
\label{sec:lum}

The total fluxes measured from all point sources detected at both 3.6
and 4.5~\micron{} are listed in Table~\ref{tab:detstats}. To avoid
including bright foreground stars, sources brighter than an apparent
magnitudes of $m_{3.6} = 14$~mag are excluded from the total flux
unless they are optically identified as an AGB or RSG star or have
colors $[3.6] - [4.5] > 0.3$. The 3.6~\micron{} flux values quoted
here tend to be higher than the integrated 4.5~\micron{} fluxes
presented in \citet{lee06} and \citet{jackson06}, which include only
the optical and near-IR extent of each galaxy instead of the entire
3.6 and 4.5~\micron{} IRAC coverage.  The flux estimates presented
here are, of course, overestimates, as they do not exclude the
contribution from background galaxies.

In Table~\ref{tab:detstats}, we also compare the summed flux of
sources one magnitude above the TRGB to that of one magnitude below
the TRGB. Choosing to compare the relative fluxes within narrow
magnitude intervals directly above and below the TRGB makes this
approach less susceptible to the effects of magnitude-dependent
completeness and contamination (from galaxies). We find that, in
general, this ratio is higher for galaxies with more recent star
formation. This correlation confirms that older stellar populations
have fewer high-luminosity stars. DDO~210 and Sextans~A do not follow
the trend, but this is due to incompleteness within one magnitude
below the TRGB, since these galaxies are the most distant. LGS~3 is
the only other galaxy that does not follow the trend; this may be due
to a combination of photometric incompleteness and low number counting
statistics.

\subsection{\emph{Derived Stellar Mass}}
\label{sec:mass}

As discussed in the previous section, the number of super-TRGB stars
decreases as a galaxy ages, so that the older the stellar population,
the higher the underlying stellar mass represented by each super-TRGB
star. Following this premise, the \citet{vanloon05} prescription for
determining the total stellar mass of a galaxy based on the number of
stars brighter than the TRGB in the {\it L\arcmin{}}-band
(3.76~\micron{}) yields stellar masses ranging from $5.5 \times 10^5 -
1.2 \times 10^7 M_\odot$ (Table~\ref{tab:compfrac}), assuming an age
of 2~Gyr for the current super-TRGB population and excluding estimates
of the numbers of background galaxies from S-COSMOS and foreground
stars from \citet{robin03}. These masses are uncertain because they
are determined using a single age population and because of incomplete
sky coverage in some galaxies. If we adjust the age to coincide with
past star formation events (see \S\,\ref{sec:compare}), we find a
wider range of stellar masses (Table~\ref{tab:compfrac}). Because of
the assumption of a single-aged stellar population, our mass
estimates, with the exception of the estimates for Pegasus~dIrr and
WLM, are somewhat larger than, but are still within reasonable
agreement to, those found by \citet{lee06} from the
mass-to-metallicity relation.

\section{THE AGB STARS}
\label{sec:agbs}

It is important to determine if photometric incompleteness affects our
AGB statistics. Photometry in Phoenix, LGS~3, DDO~210, Leo~A, and
Pegasus~dIrr is 50\% complete to more than one magnitude fainter than
the TRGB for the reddest sources.  The location of the Sextans~A TRGB
is unknown, but assuming it is near $M_{3.6} = -6.1$~mag (dotted line
in Fig.~\ref{fig:lfunc}), the Sextans~A 50\% completeness limit is
$\approx$0.4~mag fainter than the TRGB for the reddest sources.  The
bluest sources are 50\% complete down to $\gtrsim 0.3$~mag less than
the TRGB in Phoenix, LGS~3, DDO~210, Leo~A, and
Pegasus~dIrr. Photometry of the bluest sources in Sextans~A is not
complete down to $M_{3.6} = -6.1$~mag.  Therefore, without considering
the very small number of objects not detected due to crowding, our
data represent a complete census of super-TRGB stars in the central
5.8\arcmin{} $\times$ 5.8\arcmin{} regions of the sampled galaxies,
with the exception of the bluest super-TRGB stars in Sextans~A. Note
that because the least massive AGBs are also the least luminous,
barely reaching luminosities exceeding the TRGB \citep{mcdonald08},
we have underrepresented AGB stars from the first epoch of star
formation.

\subsection{\emph {Optical Completeness}}
\label{sec:complete}

After subtracting background galaxies (from S-COSMOS), we detect 37 --
478 IR AGB candidates in the six galaxies discussed in this paper
(Table~\ref{tab:compfrac}).  Using the same criteria for defining an
AGB candidate, which are slightly different from the criteria used in
papers I and II, IC~1613 and WLM contain 594 and 451 AGB candidates,
respectively. Of these candidates, 35\% -- 100\% are detected in the
optical, {\it above the I-band TRGB}.  While larger percentages are
detected in the optical below the {\it I}-band TRGB, these sources
would have been misidentified in the optical and excluded from AGB
star studies, thus we define the optical completeness fraction as the
fraction of AGB candidates in the IR that are correctly identified in
the optical (Table~\ref{tab:compfrac}). The equivalent percentages in
IC~1613 and WLM are not presented in papers~I and II, but are computed
here to be 79\% and 63\%, respectively.

\begin{figure}[h!]
\epsscale{1} \plotone{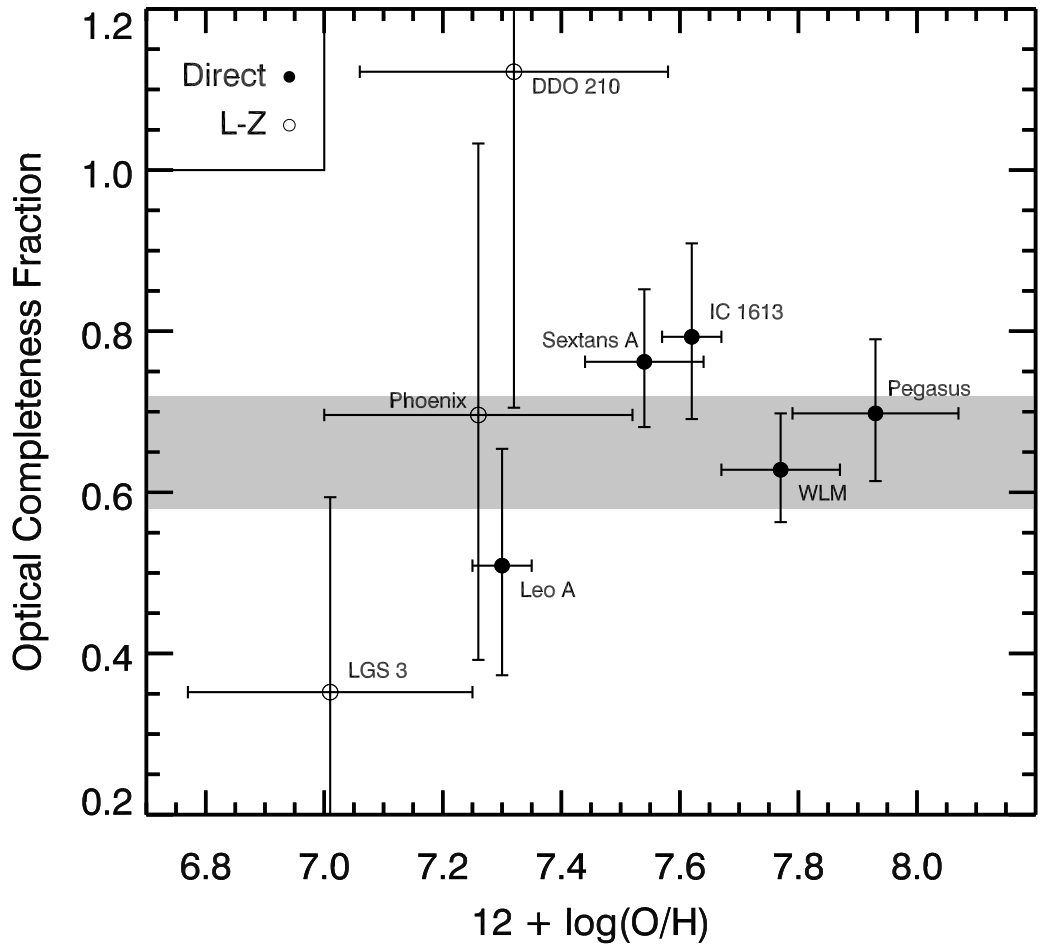}
\figcaption{Optical completeness fraction vs. metallicity. Closed
circles are galaxies with metallicities from \citet{vanzee06}, and
open circles show galaxies whose metallicities were determined using
the $L$-$Z$ relationship from \citet{lee06}. The optical completeness
fraction is computed by first estimating the true background galaxy
contamination by fitting a flat contribution to the radial density
profile of IR AGB candidates (Fig.~\ref{fig:profiles}). The errors
include the 1\,$\sigma$ uncertainties in the fit, Poisson statistics,
and foreground star contamination. We find no convincing correlation
between optical completeness and metallicity, indicating that dust
production is not inhibited at low metallicity. An optical
completeness fraction of approximately 60\% to 70\% is consistent with
the data for all eight galaxies (shaded
region).\label{fig:metallicity}}
\end{figure}

Figure~\ref{fig:metallicity} shows the metallicity versus optical
completeness fractions after correcting for background contamination,
as determined from fitting the IR radial profiles
(\S\,\ref{sec:bkgnd}). The error bars reflect the fitting
uncertainties, which are largely due to the small fields of view.  We
have adopted 12\,+\,log(O/H) values from \citet{vanzee06}, and used
the luminosity-metallicity ($L$-$Z$) relationship from \citet{lee06}
to estimate 12\,+\,log(O/H) for Phoenix, DDO~210, and LGS~3, all of
which have no \ion{H}{2} regions. No trend with metallicity is
apparent in Figure~\ref{fig:metallicity}, and it is clear that an
optical completeness fraction of roughly 60\% --70\% is consistent
within the errors in all eight galaxies. The AGB in optical surveys is
therefore underrepresented by $\approx$35\%, which may have important
consequences for SFHs derived from isochrone fitting to optical CMDs.

\begin{figure}[h!]
\epsscale{1} \plotone{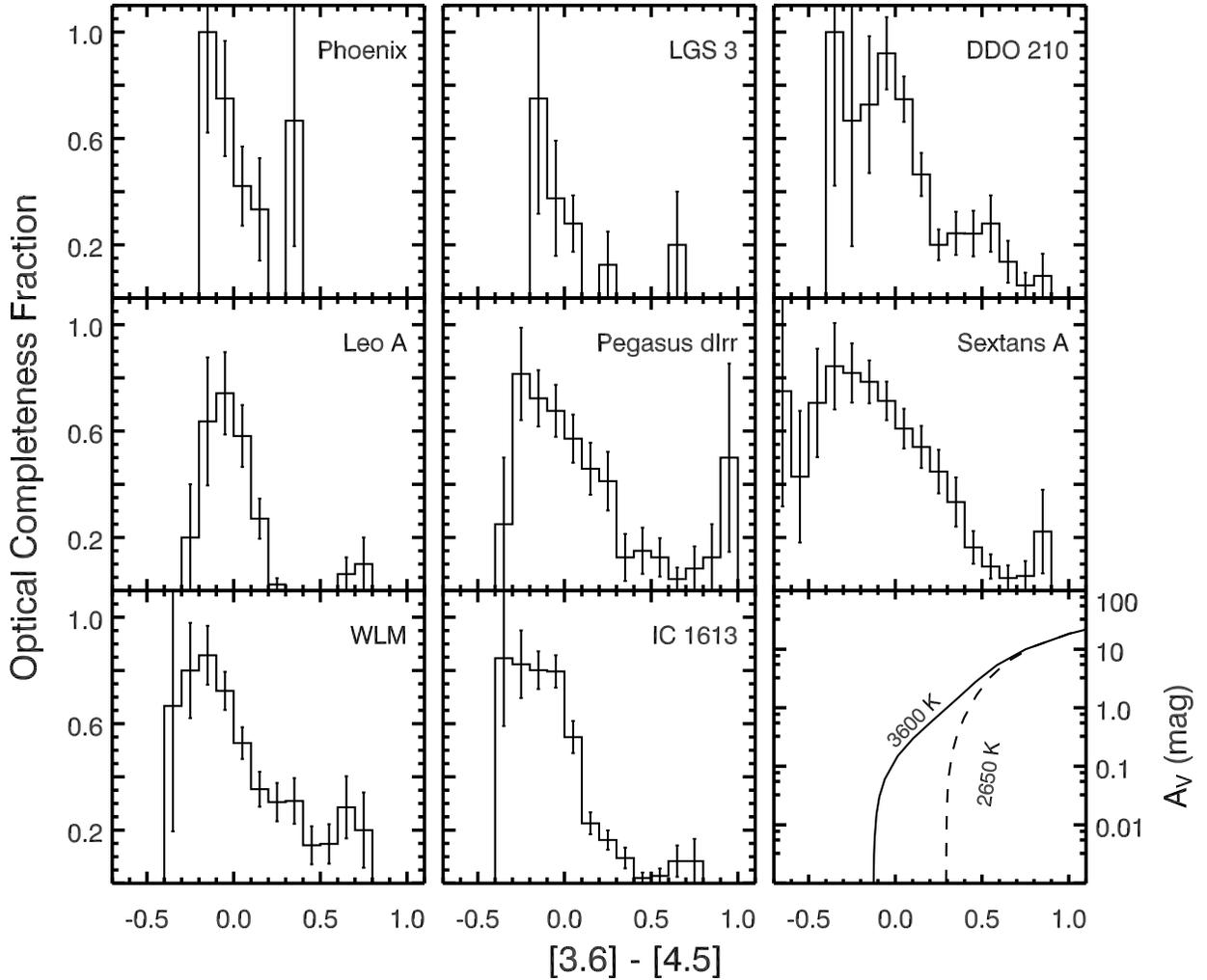} \figcaption{Fraction of
  stars detected in the optical as a function of $[3.6] - [4.5]$.  The
  optical completeness fraction includes only stars detected above the
  3.6~\micron{} TRGB {\it and} the {\it I}-band TRGB. The errors bars
  were determined by taking the square root of the number of optically
  detected AGB stars and dividing by the total number of IR-identified
  AGB candidates and therefore only represent the degree to which
  small-number statistics might affect the trend. The lower right
  panel shows the visual extinction as a function of $[3.6] - [4.5]$
  for a dust composition of 85\% AMC and 15\% SiC and effective
  temperatures of 3600 K (solid line) and 2650 K (dashed line). In all
  eight galaxies, the optical completeness decreases as a function of
  optical depth, as expected, reaching zero near $[3.6] - [4.5]
  \approx 0.5$ in most cases.  In all galaxies but Sextans~A and WLM,
  there is a secondary dip in completeness near a color of 0.2, which
  may be due to the presence of background galaxies and indicates
  that, in these galaxies, there are very few AGB stars redward of
  this drop-off.\label{fig:compfrac}}
\end{figure}

In Figure~\ref{fig:compfrac}, we show the fraction of optical
completeness (as defined in this paper) as a function of $[3.6] -
[4.5]$ color.  The optical extinction ($A_V$) is shown in the lower
right panel. The optical completeness declines as a function of color,
approaching 0\% for the reddest sources, as expected if the
incompleteness is due to circumstellar extinction.  For DDO~210,
Pegasus~dIrr, Sextans~A, WLM, and IC~1613, the completeness fraction
approaches zero roughly between $0.5~<~[3.6]~-~[4.5]~<~1.0$.  In
Phoenix, LGS~3, and Leo~A, the completeness approaches 0\% at bluer
colors ($[3.6] - [4.5] \approx 0.25$), which is due to the small
populations of very dusty AGB stars and to strong background galaxy
contamination in these galaxies. Background galaxies are
particularly red objects and thus drop out of the optical at fairly
blue colors compared to dusty AGB stars.

As discussed in \S\,\ref{sec:bkgnd}, the IR AGB candidate populations
are heavily contaminated by background galaxies, and the same trend of
decreasing optical completeness with IR color is also consistent with
this contamination. Based on fits to the stellar radial profiles,
Pegasus~dIrr, Sextans~A, and WLM have the smallest percentage
background galaxy contamination in the IR AGB candidate population. It
is interesting to note that two of these three low-contamination
galaxies are the only two galaxies in our sample that do not show a
sudden drop in completeness near $0.1 < [3.6] - [4.5] < 0.2$, and in
the case of Pegasus~dIrr, the drop occurs at an intermediate color
($[3.6] - [4.5] \approx 0.3$). This may suggest that at the $[3.6] -
[4.5] = 0.1 - 0.2$ color transition, background galaxies become the
dominate source type. In galaxies that show a drop in completeness
beyond this color transition, there may be few, if any, {\it very red}
AGB stars.

\begin{figure}[h!]
\epsscale{1} \plotone{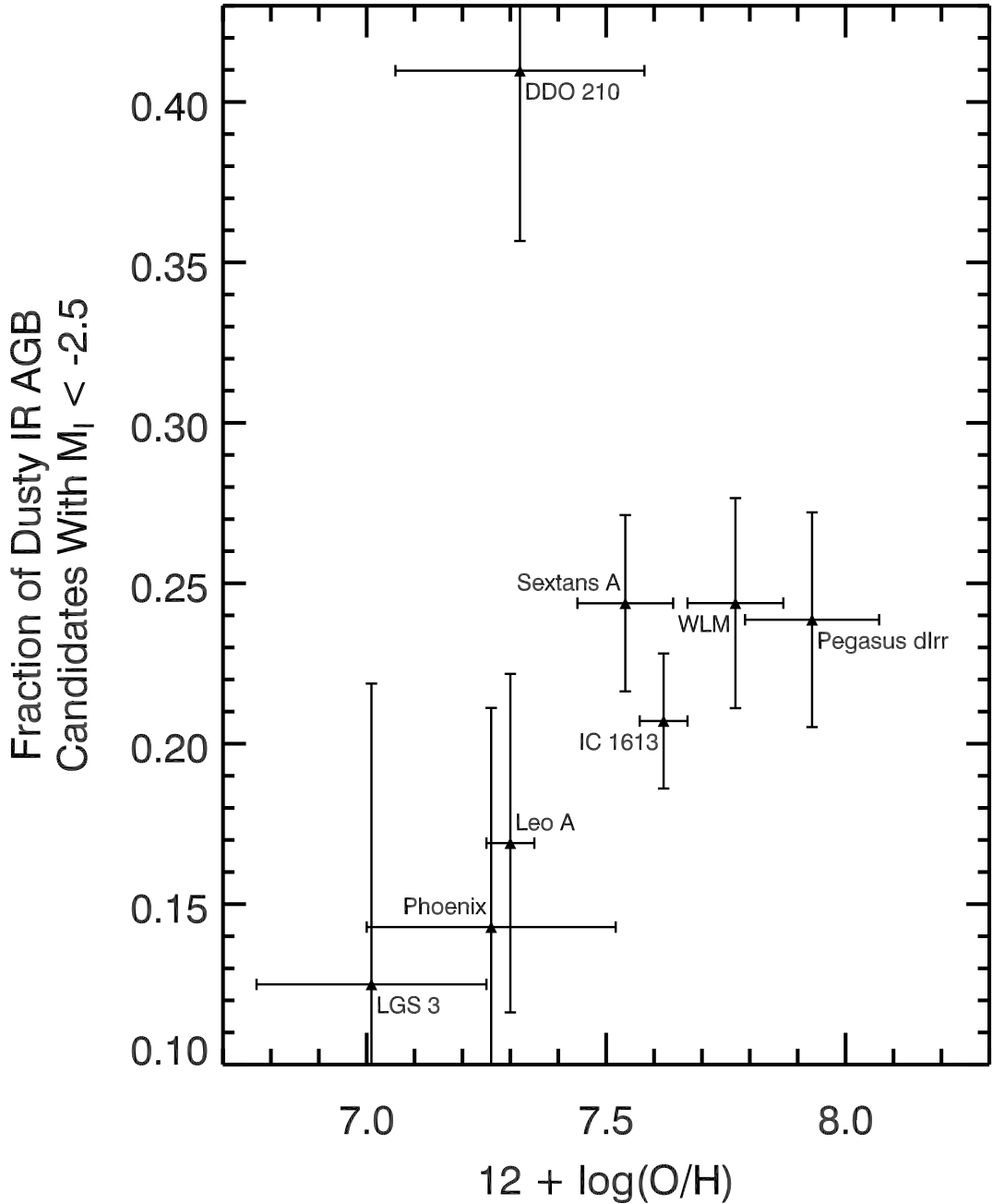} \figcaption{Fraction of red
  stars in each galaxy.  The fraction includes only IR AGB candidates
  that are detected in the optical with $M_{\rm I} < -2.5$~mag, and
  therefore does not include the most extreme dusty stars. A source is
  considered ``red'' if $[3.6] - [4.5] > 0.2$.  With the
  exception of DDO~210, we see a clear trend for higher metallicity
  galaxies to have a higher fraction of moderately red stars. This
  trend may be due to metallicity, but may also be due to the age of
  the populations.
  \label{fig:dusty}}
\end{figure}

Figure~\ref{fig:dusty} shows the fraction of IR AGB candidates {\it
with} optical counterparts (i.e., {\it not} background galaxies) that
have $[3.6] - [4.5] > 0.2$. The trend in Figure~\ref{fig:dusty}
suggests a scarcity of very red AGB stars in some galaxies in our
sample. With the exception of DDO~210, there is a clear correlation
between metallicity and the fraction of red stars, as expected if dust
is produced more efficiently in environments that are rich in metals.

An alternate interpretation of Figure~\ref{fig:dusty} is that stellar
populations older than $\approx 3.0$~Gyr are unable to produce many
stars with $[3.6] - [4.5] > 0.2$, regardless of metallicity
\citep{marigo08}. The galaxies with the smallest fraction of their
current stellar population formed within the last 3~Gyr
\citep{orban08} are indeed the galaxies with the fewest very red
stars, with the exception of DDO~210.

It is generally assumed that lower metallicity systems will be less
dusty than their higher-metallicity counterparts, resulting in a
smaller population of dust-enshrouded AGB stars.  This is indeed what
we see in Figure~\ref{fig:dusty}. This should lead to a trend of
decreasing optical completeness with increasing metallicity.
Nevertheless, we see no convincing correlation between optical
completeness and metallicity in Figure~\ref{fig:metallicity},
suggesting that even environments with very low metallicity \citep[$Z
< 2\%~Z_\odot$, assuming $Z_{\odot} = 0.0122$;][]{asplund04} are
capable of forming at least a few heavily enshrouded stars. This is
the case even if the majority of the stars in these galaxies harbor
little to no dust at all, either due to low metallicity or to an older
stellar population.

This result supports an increasing number of observations suggesting
that dust formation is not prohibited at low metallicity
\citep{cannon02, boyer06, gruendl08, sloan08, lagadec08}, although it
is forming in smaller quantities \citep{boyer06, mcdonald07,
vanloon08}. It may be that the ability to form dust has little or no
dependence on the initial metal-content, rather, nucleosynthesis in
the stars themselves and dredge-up in radial-pulsing AGB stars is the
dominant source of dust-forming metals in the circumstellar
envelope. However, we must emphasize that we do not have metallicity
measurements for any individual stars in our sample, and it is
therefore possible that the reddest (i.e., dustiest) stars have higher
metallicities than the bulk population.

\subsection{\emph {The Carbon Stars}}
\label{sec:carbon}

In addition to the carbon stars in WLM and IC~1613 (Papers~I and II),
populations of carbon stars are known in three of the galaxies studied
here: DDO~210, Phoenix, and Pegasus~dIrr. The remaining three galaxies
may also contain carbon stars, but to our knowledge no searches have
yet been carried out. Given the rough estimates of the C/M ratios
below, we might expect at least a handful of carbon stars in LGS~3,
Leo~A, and Sextans~A. Although, since carbon stars require a mass
larger than 1~$M_\odot$ to form, we might expect few, if any, carbon
stars in metal-poor, low star forming galaxies like LGS~3.

\subsubsection{Phoenix}
\label{sec:phoenix-carbon}

\begin{figure}[h!]
\epsscale{1} \plotone{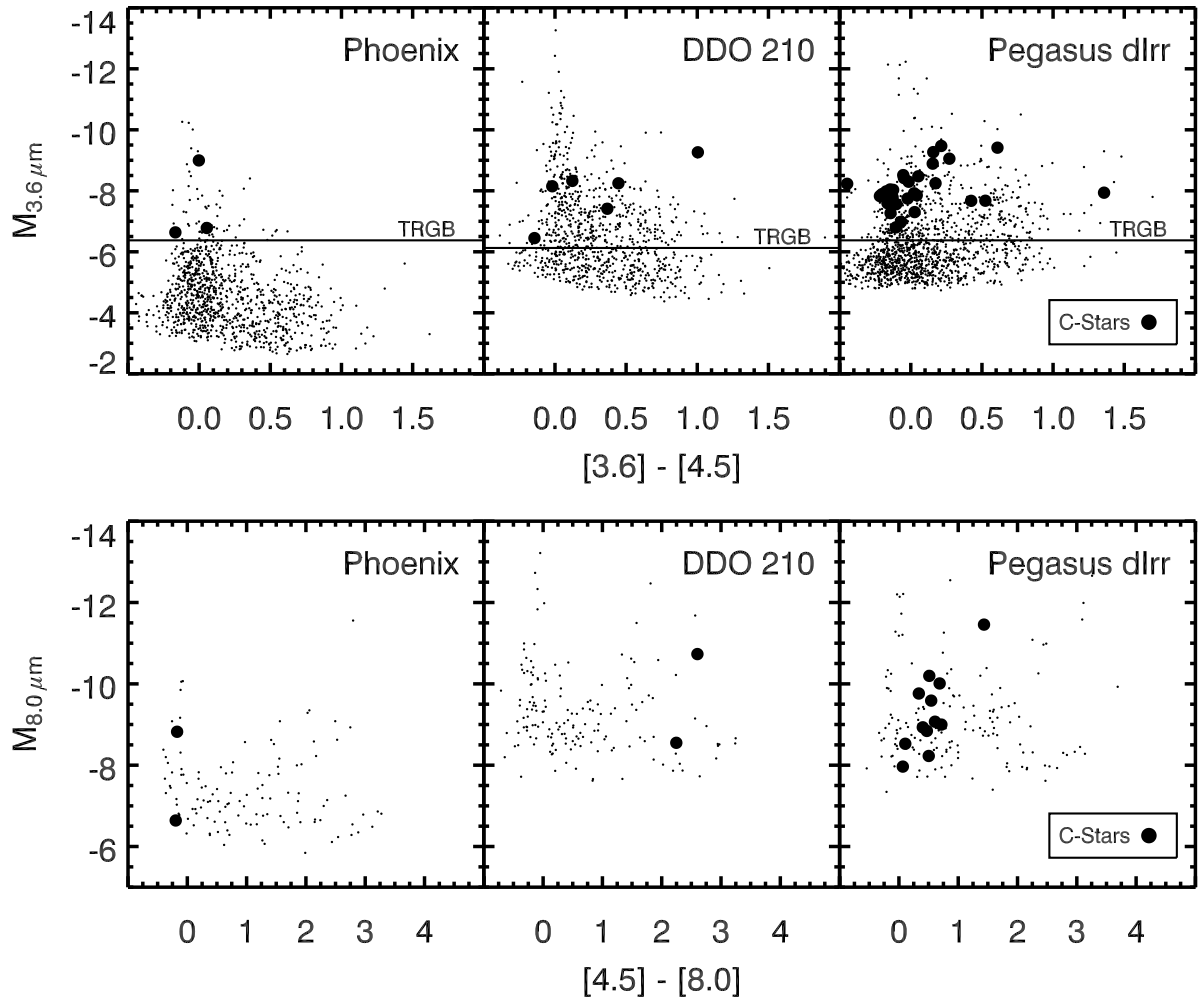} \figcaption{Carbon stars on
  the $[3.6]$ vs. $[3.6] - [4.5]$ and $[8.0]$ vs. $[4.5] - [8.0]$
  color magnitude diagrams. The small black dots mark all of the IRAC
  detections, and the large black circles mark carbon stars identified
  in Phoenix \citep{dacosta94,menzies08}, DDO~210
  \citep{battinelli00}, and Pegasus~dIrr \citep{battinelli00}. The carbon
  stars with $[3.6] - [4.5] > 0.2$ are likely losing significant
  mass. \label{fig:cstars}}
\end{figure}

Phoenix appears to have a very small population of three carbon stars
\citep{dacosta94,menzies08}. These three stars were identified through
spectroscopic measurements. The brightest carbon star
(Fig.~\ref{fig:cstars}) is a Mira variable with a period of 425 days
\citep{menzies08}.  All three carbon stars in Phoenix have relatively
blue $[3.6] - [4.5]$ colors, which may be due to the low metallicity
of the galaxy causing a low optical depth in the circumstellar
envelope. If all three stars belong to Phoenix, and the remaining 28
IR-identified AGB candidates identified in this work are M-stars, then
the C/M ratio for Phoenix is a mere 0.11.

\subsubsection{DDO~210}
\label{sec:ddo210-carbon}

Using the CN/TiO technique with narrowband optical photometry,
\citet{battinelli00} found 3 carbon stars in DDO~210, two of which are
located within the {\it Spitzer} coverage and are detected in
IRAC. With broadband near-IR photometry, \citet{gullieuszik07} found
an additional 6 carbon stars, 4 of which are located within our {\it
Spitzer} coverage and are detected in IRAC. The two
\citet{gullieuszik07} stars with the reddest $[3.6] - [4.5]$ colors in
Figure~\ref{fig:cstars} appear to be Long Period Variables, typical of
mass-losing carbon stars.  One other carbon star is quite red, while
the remaining three have relatively blue colors, similar to the carbon
stars in Phoenix.

\citet{battinelli00} also find 158 M-stars in an 8.2\arcmin{} $\times$
8.2\arcmin{} field. By measuring the density of M-stars on the
outskirts of their CCD image, \citet{battinelli00} claim that all 158
M-stars are in the foreground, however a later estimate of the C/M
ratio in \citet{battinelli05} using modified a distance and reddening
computes a C/M ratio of 0.09 $\pm$ 0.08.  Assuming an even distribution of all
158 M-stars, we expect 79 of these stars to fall within our IRAC
coverage. In \S\,\ref{sec:foreground}, we found that only 32
foreground stars are expected in the area covered by our IRAC
images. If \citet{battinelli00} overestimated the number of foreground
M-stars, we are left with 47 M-stars and 6 carbon stars covered in our
IRAC images, yielding a C/M ratio of 0.13.  If all of the
IR-identified AGB candidates (minus background and foreground
contamination) reported here (Table~\ref{tab:detstats}) are truly AGB
stars belonging to DDO~210, then only 26\% of the total IR AGB star
population is used to compute this ratio, and it should not be
over-interpreted.

\subsubsection{Pegasus~dIrr}
\label{sec:ddo216-carbon}

The population of carbon stars in Pegasus~dIrr is comparable to the
populations found in IC~1613 and WLM.  With optical narrowband
observations, \citet{battinelli00} find 40 carbon stars in
Pegasus~dIrr, 35 of which fall inside our IRAC coverage, and 32 of
which are detected here (three are undetected due to crowding).  The
bulk of the carbon stars have relatively blue $[3.6] - [4.5]$ colors
(Fig.~\ref{fig:cstars}); four are very red, and five are very bright
at 3.6~\micron{}. Pegasus~dIrr has the highest metallicity of the
galaxies in our sample (12\,$+$\,log(O/H) = 7.93), so it seems less
likely that the colors are blue entirely because of a low optical
depth, as may be the case for Phoenix and DDO~210. Comparison to
Figure~\ref{fig:cmd+iso} shows that the brightest carbon stars in
Pegasus~dIrr match the 560~Myr population better than the older and
younger isochrones, which agrees with the enhanced star formation that
occurred just short of 1~Gyr ago \citep{dolphin05}.

Assuming a color excess of E(B--V) = 0.03, the C/M ratio quoted by
\citet{battinelli00} is 0.73, and uses a total of 175 AGB stars in an
area roughly twice the size of our IRAC coverage. However, based on
the numbers of C-stars and M-stars quoted in Table~6 of
\citet{battinelli00}, we cannot reproduce the C/M ratio and suspect a
typographical error; with their numbers, we recalculate C/M $=$
0.50.\footnote{Private communication with S. Demers has confirmed this
recalculation.}  A later modification of the distance and reddening
towards Pegasus~dIrr adjusts the C/M ratio to 0.62 $\pm$ 0.22
\citep{battinelli05}. According to the population synthesis models of
\citet{robin03} (see \S\,\ref{sec:foreground}), \citet{battinelli00}
overestimated the contamination of M-stars in the foreground by a
factor of $\approx$2.8. Therefore, assuming the population synthesis
model is correct and that the distribution of M-stars is flat,
yielding 58 M-stars in the IRAC coverage, we recompute the C/M ratio
to be 0.60, well within agreement to the most recent C/M ratio
calculated by \citet{battinelli05}. This ratio is computed with a
total of 93 AGB stars, which is only 34\% of the IR AGB candidates
reported here.

\subsection{\emph {Mass-Loss Rates}}
\label{sec:mlrs}

As in papers~I and II, we have estimated the MLRs of the AGB candidates in
each galaxy by comparing the IR fluxes to the radiative transfer
models of \citet{groenewegen06}. We linearly interpolated the $[3.6] -
[4.5]$ colors onto the \citet{groenewegen06} models to estimate the
optical depth ($\tau$), which corresponds to a particular IR color
given assumptions about the wind composition and stellar effective
temperature ($T_{\rm eff}$). The resulting MLRs were then scaled
according to \citet{vanloon06}, where $\dot{M} \propto
\tau\psi^{-(1/2)}L^{(3/4)}$, $\psi$ is the dust-to-gas ratio, and $L$
is the stellar luminosity.  The dust-to-gas ratio scales as $\psi =
\psi_{\odot}10^{[Fe/H]}$ and $\psi_{\odot} = 0.005$
\citep{vanloon05}. See Table~\ref{tab:avgmlr} for the average stellar
MLRs and Table~\ref{tab:cummlr} for integrated galaxy MLRs.

The integrated and average MLRs computed using the entire population
of IR AGB candidates are most certainly overestimates, as 21\% -- 67\%
of these populations are background galaxies. For this reason, we have
also computed the integrated and average MLRs for only sources that
have optical counterparts with $M_{\rm I} < -2.5$~mag. This approach
excludes any AGB stars that are obscured in the optical, and thus
excludes the stars with the most dust and highest MLRs. To regain some
of these interesting sources, we have also computed the integrated
MLRs for all sources that either have optical counterparts {\it or} are
brighter than $m_{3.6} = 16$~mag, the approximate cut-off between
unresolved background galaxies and AGB stars.  These totals are listed
in Table~\ref{tab:cummlr}.  We have also computed the integrated MLR
by subtracting the contribution from the number of galaxies predicted
in the S-COSMOS data, assuming each source has the average MLR. This
approach inserts more uncertainty into the resulting integrated MLR
since background galaxies can artificially inflate or deflate the
average, depending on their (incorrectly-assigned) absolute
magnitudes.

\begin{figure}[h!]
\epsscale{1} \plotone{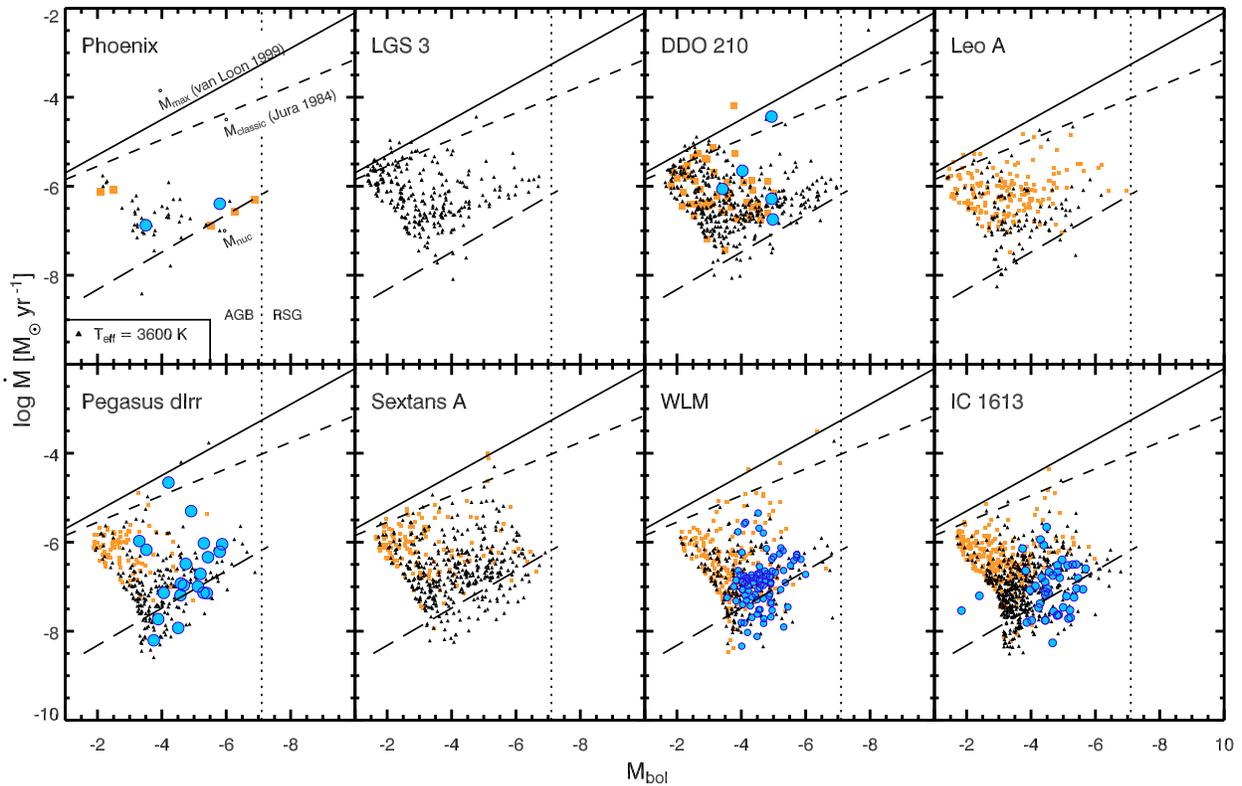} \figcaption{Mass-loss
rates vs. bolometric magnitudes for all objects brighter than the
3.6~\micron{} TRGB, fainter than the AGB limit, and {\it not}
optically identified as an RSG or a blue object.  Orange squares are
sources with $M_{\rm I} > -2.5$~mag and $m_{3.6} > 16$~mag and are heavily
contaminated with background galaxies. Large blue circles are carbon
stars.  There are no Orange squares in LGS~3 due to the difficulty in
matching IRAC sources to ACS sources (\S\,\ref{sec:optphot}). Mass-loss rates were derived
using the \citet{groenewegen06} models (\S\,\ref{sec:mlrs}), assuming
a wind composition of 85\% AMC and 15\% SiC and an effective
temperature of $T_{\rm eff} = 3600$ K. The short-dashed line marks the
classical single-scattering limit from \citet{jura84} ($\dot{M}_{\rm
classic}$) and the solid line marks the maximum mass-loss rates
($\dot{M}_{\rm max}$) observed in the Large Magellanic Cloud
\citep{vanloon99}.  The lower, long-dashed line marks the
nuclear-burning mass consumption rate ($\dot{M}_{\rm nuc}$) and the
vertical, dotted line marks the AGB limit at $M_{\rm bol} =
-7.1$~mag. Several stars in each galaxy, save Phoenix, are above the
single-scattering limit and have large optical
depths.\label{fig:massloss}}
\end{figure}

The stellar MLR versus bolometric magnitude for each galaxy is shown
in Figure~\ref{fig:massloss}, assuming a dust composition of 85\% AMC
$+$ 15\% SiC and $T_{\rm eff} = 3600$~K. We have included in this
figure only sources that fall between the 3.6~\micron{} TRGB and the
AGB limit and are not identified in the optical as an RSG or a blue
object.  Sources below the background galaxy cut-off at $m_{3.6} =
16$~mag that do not have optical counterparts are plotted in small
orange points (except for LGS~3, which is not matched to the optical
data); these points represent the population that is likely dominated
by background galaxies. Carbon stars are plotted in filled blue
circles.  We find that the most heavily mass-losing stars are among
the most luminous AGB candidates. The maximum MLRs in all eight
galaxies are in good agreement with that found in the LMC
\citep{vanloon99}, which is somewhat higher than the classical
single-scattering MLR limit predicted by \citet{jura84}, above which a
dusty circumstellar envelope results in multiple scattering of photons
\citep{gail86}. The sources that lie above the single-scattering limit
are indeed those with the largest optical depths, as determined from
their $[3.6] - [4.5]$ colors.

While the vast majority of the AGB candidates have moderate MLRs ($<
10^{-6} M_{\odot}~{\rm yr}^{-1}$), there are a handful of sources in
each galaxy, with the exception of Phoenix, that are near or above the
classical single-scattering limit \citep[short-dashed line in
Figure~\ref{fig:massloss};][]{jura84}. The brightest of these have the
strongest MLRs and dominate each galaxy's integrated MLR. A few of
these heavily mass-losing sources are also undetected in the optical
(orange points) and may prove to be background galaxies. However, in
Leo~A, Sextans~A, WLM, and IC~1613, many of these optically obscured,
heavily mass-losing sources (1, 2, 4, and 5 sources, respectively) are
brighter than the background galaxy cut-off ($m_{3.6} < 16$~mag), and are
likely true obscured AGB stars.

The majority of the AGB candidates are in the superwind phase
\citep{vanloon05,girardi07}, i.e., the MLRs exceed the nuclear-burning
mass consumption rate ($\dot{M}_{\rm nuc}$; long-dashed line in
Fig.~\ref{fig:massloss}). Mass loss is the dominant evolutionary
driver for these stars since they will expel their outer layers before
exhausting their nuclear fuel.

The timescale of the AGB superwind phase is short
\citep[$\sim$$10^6$~yr;][]{vanloon05,girardi07}. If we assume that
each galaxy hosts a broad range of AGB star masses, then we may also
assume that each AGB star loses an average total of
1.5~$M_\odot$. Using the mean individual MLRs listed in
Table~\ref{tab:avgmlr} (averaged over all five combinations of dust
composition and effective temperature), the superwind timescale we
derive ranges from $4 \times 10^5 - 2 \times 10^6$~yr, which agrees
well with previous estimates. A consequence of this short timescale is
that the integrated MLR of low-mass galaxies may be highly
time-variable, as is the case for star clusters \citep{vanloon08,
mcdonald08}, and must not be over-interpreted.  Within a few million
years, a mere handful of new, strongly mass-losing AGB stars may
appear in some galaxies, but not in others, causing a strong increase
in the integrated MLR of the former.  Therefore, we do not claim that
the integrated MLRs quoted here are representative of any particular
galaxy trait.

\subsubsection{Uncertainties in the Mass-Loss Rates}
\label{sec:mlrunc}

As emphasized in papers I \& II, there are major uncertainties in
deriving MLRs for AGB stars using only 3.6 and 4.5~\micron{}
photometry.  The first is a degeneracy between $\tau$ and $T_{\rm
eff}$: cool stars with small MLRs can have the same $[3.6] - [4.5]$
color as warmer stars with strong MLRs.  This uncertainty most
strongly affects the bluest AGB candidates($[3.6] - [4.5] < 0.5$), but
is of little consequence for the redder AGB candidates, where the MLR
is much less sensitive to $T_{\rm eff}$. This uncertainty may also be
exacerbated by atmospheric molecular absorption, which can affect IRAC
colors, although these effects have not yet been quantified
\citep{marengo07}. Since the reddest AGB stars also have the strongest
MLRs, the integrated MLR for each galaxy and the individual MLRs for
the red stars is robust.

The second major uncertainty lies with the choice of the chemical
composition of the AGB winds.  With no information about the wind
compositions, we must make the unlikely assumption of a single
composition for the entire AGB population. The integrated MLRs
computed for different wind compositions vary by up to a factor of eight.

While these uncertainties affect the MLRs of individual sources, the
integrated MLR is heavily affected by the unfortunate placement of background
galaxies in the same region of the CMD as the mass-losing AGB
stars, as discussed above. While we have attempted to remove this
contamination, we must still emphasize that the integrated MLRs
reported here are meant as first-order estimates only. 

\subsubsection{ISM Gas Return}
\label{sec:ism}

While massive stars occasionally provide bursts of feedback to the ISM
when they explode as supernovae, it is the more numerous, mass-losing
stars that dominate long-term gas return.  Gas return from massive
stars occurs only during and shortly after a burst of star formation,
whereas AGB gas return is continuous over time and allows galaxies to
gradually rebuild an ISM while remaining in quiescensce with regard to
star formation. A comparison between the the returned mass and the
mass used to form stars ($f_{\rm return} = \dot{M}_{\rm dusty} / {\rm
SFR}_{\rm (1\,Gyr)}$), computed using current integrated MLRs (bottom
panel of Table~\ref{tab:cummlr}) and the average SFRs over the last
1~Gyr \citep{orban08}, shows that each galaxy in our sample has been
undergoing a SFR that is unsustainable, with the exception of LGS~3
and DDO~210. Without another source of gas to the ISM or a change in
star formation rate, these galaxies will exhaust their supply of
interstellar gas (see Table~\ref{tab:ism}). We note that there are
uncertainties associated with converting SFHs derived from partial
fields to global SFHs, but we believe that these SFHs should be
qualitatively representative of the galaxies.

It is unlikely that these galaxies can accrete significant amounts of
gas from their surrounding environments, as they are all isolated from
other systems. If these galaxies are to maintain their current SFRs,
there must be some non-dusty mass loss in addition to the dust-traced
mass loss measured here. If the ISM is replenished solely by stellar
mass loss (dusty plus non-dusty), then $f_{\rm return}$
(Table~\ref{tab:ism}) reflects what fraction of the total stellar mass
loss is traced by dust.  We find that in Sextans~A, WLM, and IC~1613,
dusty mass loss accounts for only 2\% -- 4\% of the total mass loss
required to replenish the ISM at a rate equal to the current SFR,
while in Phoenix, Leo~A, and Pegasus~dIrr, this fraction rises to 22\%
-- 30\%. Previous work by \citet{mcdonald07} and \citet{mcdonald08}
found that roughly 50\% of the mass loss is traced by dust in globular
clusters.  The very large amount of non-dusty mass loss required to
balance the SFRs in Sextans~A, WLM, and IC~1613 suggests that the
current SFRs in these galaxies are indeed unsustainable.

The AGB stars in LGS~3 and DDO~210 are losing mass at a rate faster
than new stars are currently being formed. Therefore, barring any
outside influences, the mass loss traced by dust in these galaxies is
sufficient to replenish the ISM at a rate that can sustain the ISM
recycling rate indefinitely, eliminating the need for other gas return
mechanisms such as non-dusty mass loss and gas accretion.

By comparing the \ion{H}{1} mass in each galaxy with the current SFR
and dusty MLR, we can compute the life expectancy of each galaxy, or
the timescale before the galaxy runs out of \ion{H}{1} gas.  This
timescale is known as the ``Roberts time'' \citep{sandage86},
corrected so that it now includes gas return:

\begin{equation}
\tau_{\rm R, corrected} = \left(1 - \frac{\dot{M}_{\rm dusty}}{{\rm SFR}}\right)^{-1}~\frac{M_{\rm gas}}{{\rm SFR}}. 
\end{equation}

\noindent The correction factor is referred to as the recycling
factor. We find that Leo~A and Phoenix can survive as gas-rich dwarf
galaxies for 17~Gyr and 76 Gyr, respectively, but that Pegasus~dIrr,
Sextans~A, WLM, and IC~1613 can continue star formation activity at
the current rate for only 2.5~Gyr -- 6.8~Gyr, with WLM being the first
galaxy to become inactive (Table~\ref{tab:ism}). For Leo~A, Phoenix,
and Pegasus~dIrr, dusty mass loss is responsible for an increase of
29\% -- 40\% over the uncorrected $\tau_{\rm R}$, while in the
remaining three galaxies, dusty mass loss is responsible for a mere
2\% -- 4\% increase.  For disk galaxies, \citet{kennicutt94} finds
that this increase can be significantly larger, although we note that
those estimates include mass lost during all stages of stellar
evolution.

Assuming a closed-box model, and that the average SFR over the last
5~Gyr represents the average rate over each galaxies' lifetime, then
when including gas return from dusty stellar mass loss, we find that
the predicted gas-rich lifetimes of WLM and IC~1613 ($\tau_{\rm
lifetime} =$ 3.7 and 4.3~Gyr, respectively, see Table~\ref{tab:ism})
are shorter than the current estimate of the mean mass-weighted
stellar ages for each galaxy \citep[$\tau_{\rm age} =$ 6.7 and
7.7~Gyr, respectively;][]{orban08}. Even if we assume that dusty mass
loss accounts for only 10\% of the total mass loss, we still find
lifetimes that are too short to allow for any current star formation
in WLM and IC~1613 ($\tau_{\rm lifetime} = $5.8 and 5.7~Gyr,
respectively). This indicates that either these galaxies have been
accreting gas from the outside or that they have been very inefficient
at using up their initial gas to form stars, although it is unclear
what might cause the latter \citep{orban08}.

\section{SUMMARY OF RESULTS AND CONCLUSIONS}
\label{sec:summary}

We present part~III of a {\it Spitzer} IRAC census of AGB stars in
eight Local Group dwarf irregular galaxies: Phoenix, LGS~3, DDO~210,
Leo~A, Pegasus~dIrr, Sextans~A, WLM, and IC~1613. Stars brighter than
the 3.6~\micron{} TRGB are identified as AGB candidates, and we find
that 50\% -- 100\% of these sources are detected and identified as AGB
stars in broadband optical photometry. An optical completeness
fraction of 70\% agrees within the errors of all eight galaxies in our
sample, suggesting that optical surveys of dwarf galaxies are
underrepresenting the AGB by 30\% -- 40\%, due to extinction from
circumstellar dust.

We find no trend of decreasing optical completeness with increasing
metallicity, as would be expected assuming that dust is more
efficiently produced at high metallicities.  This suggests that dust
production is not suppressed at low metallicity, although it may form
in smaller quantities. We do however see that the fraction of moderately
red (dusty) stars that are not obscured in the optical increases in
galaxies with higher metallicity. 

Known carbon stars were identified in our IRAC data, and we find that
previous studies have used only a small fraction (26\% -- 34\%) of the
population of M and C stars to compute C/M ratios.

Using the $[3.6] - [4.5]$ color excess, we compute average individual
stellar mass-loss rates ranging from $3.1 \times 10^{-7} - 6.6 \times
10^{-6}~M_\odot~\rm{yr}^{-1}$, agreeing well with estimates of AGB
MLRs in other studies.  Mass loss traced by dust does not appear to be
inhibited at low metallicity.  

The integrated MLRs for each galaxy range from $4.4 \times 10^{-5} -
1.4 \times 10^{-3}~M_\odot~\rm{yr}^{-1}$. Normalized to the total
stellar mass, we find a range of $\dot{M}_{\rm Tot}/M_* = 2 \times
10^{-8}$ for LGS~3, to $\dot{M}_{\rm Tot}/M_* = 6 \times 10^{-10}$ for
Leo~A. These rates do not include mass loss that is unaccompanied by
dust.

A comparison of the recent (within 1~Gyr) SFR and the current MLRs
reveals that all but two of our galaxies (LGS~3 and DDO~210) cannot
sustain their current SFR solely through ISM gas return from the
stellar mass-loss rates measured here. Phoenix, Leo~A, Pegasus~dIrr,
Sextans~A, WLM, and IC~1613 will all eventually extinguish their gas
supply unless there is an additional supply of gas.  In these
galaxies, mass loss traced by dust can account for only 2\% -- 30\% of
the material required to maintain the current SFR.

\acknowledgments We thank Alan McConnachie and Vladas Vansevi\v{c}ius
for sharing the Subaru photometry of DDO~210 and Leo~A, respectively,
and Sebastian Hidalgo and the LCID team for providing ACS data of LGS
3. We also thank Dale Jackson for many helpful discussions. Support
for this work was provided by NASA through contract 1314733, 1256406,
and 1215746 issued by JPL, Caltech to the University of Minnesota. MLB
recognizes support from the University of Minnesota Louise T. Dosdall
and Dissertation Fellowships.


\clearpage


\clearpage

\clearpage

\clearpage

\clearpage

\clearpage

\clearpage

\clearpage

\clearpage

\clearpage

\clearpage

\clearpage

\clearpage

\clearpage

\clearpage

\clearpage

\clearpage

\clearpage


\begin{deluxetable}{lcccccccc}
\tablewidth{0pc} 
\tabletypesize{\small}
\rotate
\tablecolumns{9} 
\tablecaption{Basic Galaxy Properties\label{tab:properties}}

\tablehead{
\colhead{Quantity}&
\colhead{Phoenix} &\colhead{LGS~3} &\colhead{DDO~210}&\colhead{Leo~A}
& \colhead{Pegasus~dIrr} & \colhead{Sextans~A} & \colhead{WLM} &
\colhead{IC~1613}}

\startdata
Right Ascension&&&&&&&&\\
\hspace{1em}$\alpha$(J2000)\dotfill&01$^{\mbox{\scriptsize{h}}}$51$^{\mbox{\scriptsize{m}}}$06$^{\mbox{\scriptsize{s}}}$&01$^{\mbox{\scriptsize{h}}}$03$^{\mbox{\scriptsize{m}}}$57$^{\mbox{\scriptsize{s}}}$&20$^{\mbox{\scriptsize{h}}}$46$^{\mbox{\scriptsize{m}}}$52$^{\mbox{\scriptsize{s}}}$&09$^{\mbox{\scriptsize{h}}}$59$^{\mbox{\scriptsize{m}}}$27$^{\mbox{\scriptsize{s}}}$&23$^{\mbox{\scriptsize{h}}}$28$^{\mbox{\scriptsize{m}}}$36$^{\mbox{\scriptsize{s}}}$&10$^{\mbox{\scriptsize{h}}}$11$^{\mbox{\scriptsize{m}}}$01$^{\mbox{\scriptsize{s}}}$&00$^{\mbox{\scriptsize{h}}}$01$^{\mbox{\scriptsize{m}}}$58$^{\mbox{\scriptsize{s}}}$&01$^{\mbox{\scriptsize{h}}}$0$^{\mbox{\scriptsize{m}}}$454$^{\mbox{\scriptsize{s}}}$\\

\vspace{-2mm}\\

Declination&&&&&&&&\\
\hspace{1em}$\delta$(J2000)\dotfill&$-$44\degr26\farcm7&$+$21\degr53\farcm7&$-$12\degr50\farcm9&$+$30\degr44\farcm8&$+$14\degr44\farcm6&$-$04\degr41\farcm5&$-$15\degr27\farcm8&$+$02\degr08\farcm0\\

\vspace{-2mm}\\

($m -M$)$_0$&&&&&&&&\\
\hspace{1em}(mag)\tablenotemark{a}\dotfill&$23.24\pm 0.12$&$24.54 \pm
0.15$&$25.15 \pm 0.08$ [8]&$24.51 \pm 0.12$ [3]&$24.9 \pm 0.1$&$25.61 \pm
0.07$ [4]&$24.81 \pm 08$ [11]&$24.32 \pm 0.06$ [2]\\ 

\vspace{-2mm}\\

Morphological&&&&&&&&\\
\hspace{1em}Type\dotfill&dIrr/dSph&dIrr/dSph&dIrr/dSph&dIrr&dIrr/dSph&dIrr&IrrIV-V&IrrV\\

\vspace{-2mm}\\

$M_V$&&&&&&&&\\
\hspace{1em}(mag)\dotfill&$-$10.1&$-$10.35 [9]&$-$10.6 [10]&$-$11.4&$-$12.9&$-$14.6&-14.5&-14.7\\

\vspace{-2mm}\\

12 $+$ log (O/H)\tablenotemark{b}&&&&&&&&\\
\hspace{1em}\dotfill&$7.26 \pm 0.26$&$7.01 \pm 0.24$&$7.32 \pm 0.26$&$7.30 \pm 0.05$&$7.93 \pm 0.14$&$7.54 \pm 0.10$&$7.83 \pm 0.06$&$7.62 \pm 0.05$\\

\vspace{-2mm}\\

Holmberg semiaxes&&&&&&&&\\
\hspace{1em}$a_{\rm H}, b_{\rm H}$\dotfill &\nodata &
\nodata&0.9\arcmin, 1.6\arcmin&2.3\arcmin, 3.5\arcmin&2.3\arcmin,
3.9\arcmin&3.2\arcmin, 4.0\arcmin&2.2\arcmin, 5.5\arcmin&\nodata\\

\vspace{-2mm}\\

log($M_{\rm H\,I}$)&&&&&&&&\\
\hspace{1em}($M_\odot$)\dotfill&6.5 [14]&5.0 [12]&5.3 [12]&7.0
       [14]&6.5 [5]&7.9 [1]&7.7 [6]&7.8 [13]\\

\vspace{-2mm}\\

log($M_{\rm Dynamical}$)&&&&&&&&\\
\hspace{1em}($M_\odot$)\dotfill&7.5&7.1&6.7&7.0&7.8&8.6&8.2&8.9\\

\vspace{-2mm}\\
\tablebreak
log($M_*$)&&&&&&&&\\
\hspace{1em}($M_\odot$)\tablenotemark{c}\dotfill&5.40&4.56&5.60&5.89&6.98&6.24&6.88&6.82\\

\vspace{-2mm}\\

TRGB$_{3.6~\mu\rm m}$&&&&&&&&\\
\hspace{1em}(mag)\tablenotemark{d} \dotfill&$-6.38 \pm 0.25$&$-5.88
\pm 0.25$&$-6.13 \pm 0.25$&$-6.88 \pm 0.25$&$-6.38 \pm 0.25$&$-6.13
\pm 0.25$\tablenotemark{e}&$-6.6 \pm 0.25$&$-6.2 \pm 0.20$\\

\enddata 

\tablecomments{ \ All values in this table are from \citet{mateo98},
unless marked otherwise in brackets.}

\tablerefs{ \ [1] \citet{barnes04}; [2] \citet{dolphin01}; [3] \citet{dolphin02}; [4]
  \citet{dolphin03a}; [5] \citet{hoffman96}; [6] \citet{huchtmeier81}; [7] \citet{mateo98}; [8]
  \citet{mcconnachie06}; [9] \citet{lee95}; [10] \citet{lee99}; [11]
  \citet{lee06}; [12] \citet{lo93};  [13] \citet{volders61}; [14] \citet{young96}.}

\tablenotetext{a}{ \ Distance modulus.}
  
\tablenotetext{b}{ \ Values of 12 + log(O/H) for galaxies containing
  \ion{H}{2} regions (Leo~A, Pegasus~dIrr, and Sextans~A) are taken
  from \citet{vanzee06}. For galaxies without \ion{H}{2} regions
  (Phoenix, LGS~3, and DDO~210), the $L$-$Z$ relationship determined
  by \citet{lee06} was used to determine 12 + log(O/H). To compute
  mass-loss rates, $Z$ was determined using $Z_\odot$ = 0.0122 and 12
  + log(O/H)$_\odot$ = 8.66 \citep{asplund04}.}
  
\tablenotetext{c}{ \ Stellar masses for Leo~A, Pegasus~dIrr, and
  Sextans~A are taken directly from \citet{lee06}. For Phoenix, LGS~3,
  and DDO~210, we used the mass-metallicity relation derived by
  \citet{lee06} with 4.5~\micron{} fluxes from \citet{jackson06}.}

\tablenotetext{d}{ \ The 3.6~\micron{} TRGBs as determined from this
  work (see \S\,\ref{sec:ircmd}), except for WLM (Paper I) and IC~1613
  (Paper II).}
  
\tablenotetext{e}{ \ The Sextans~A 3.6~\micron{} TRGB is not measured
    from the IRAC data directly, but by plotting the 3.6~\micron{}
    luminosity function of optical sub-TRGB stars (see
    \S\,\ref{sec:ircmd}).}
\end{deluxetable}

\clearpage

\begin{deluxetable}{lcllclcc}
\tablewidth{0pc}
\tabletypesize{\normalsize}
\tablecolumns{8} 
\tablecaption{Observation Details\label{tab:obs}}

\tablehead{\colhead{Galaxy}&\colhead{AOR key}&\colhead{$\alpha$}&\colhead{$\delta$}&\colhead{PID}&\colhead{Date}&\colhead{$\lambda$}&\colhead{Depth}\\\colhead{Name}&&\multicolumn{2}{c}{(J2000)}&&\colhead{(UT)}&\colhead{(\micron)}&\colhead{(s)}}

\startdata

LGS~3&5051393&01:03:52.65&$+$21:53:00.2&128  &2005 Jul 23&4.5, 8.0&968\\
&23043072&01:03:58.86&$+$21:49:57.1&40524&2007 Aug 12&3.6, 5.8&965\\

Phoenix&5052160 &01:51:05.65&$-$44:26:42.0&128  &2003 Dec 06&4.5, 8.0&968\\
&23043328&01:51:04.84&$-$44:30:02.9&40524&2007 Sep 07&3.6, 5.8&965\\

Leo~A&5052416 &09:59:26.23&$+$30:44:47.8&128  &2003 Dec 06&4.5, 8.0&968\\
&23042816&09:59:26.63&$+$30:44:47.5&40524&2007 Dec 28&3.6, 5.8&965\\

Sextans~A&5053696&10:11:06.45&$-$04:38:27.9&128  &2003 Dec 06&4.5, 8.0&968\\
&15892224&10:10:55.20&$-$04:44:38.6&128  &2005 Dec 24&3.6, 5.8&858\\

DDO~210&5054976 &20:46:51.70&$-$12:50:47.0&128  &2004 Oct 09&4.5, 8.0&968\\
&23043328&20:46:52.01&$-$12:50:51.2&40524&2007 Nov 13&3.6, 5.8&965\\

Pegasus~dIrr&5055744 &23:28:36.16&$+$14:44:35.1&128  &2004 Jul 26&4.5, 8.0&968\\
&23043840&23:28:44.24&$+$14:41:49.6&40524&2007 Aug 16&3.6, 5.8&965\\

\enddata
\end{deluxetable}

\clearpage

\begin{deluxetable}{llcccc}
\tablewidth{0pc}
\tabletypesize{\normalsize}
\tablecolumns{6}
\tablecaption{Sample Table: IRAC Point-Source Magnitudes\label{tab:photometry}}

\tablehead{\colhead{Galaxy} & \colhead{Source ID} &\multicolumn{4}{c}{Apparent Magnitude} \\  && \colhead{$3.6~\mu \rm m$} & \colhead{$4.5~\mu \rm m$} & \colhead{$5.8~\mu \rm m$} & \colhead{$8.0~\mu \rm m$}}

\startdata

Phoenix&  SSTU J015043.49$-$442657.4&      \nodata& 18.7$\pm$0.2&      \nodata& 16.6$\pm$0.2\\
Phoenix&  SSTU J015044.85$-$442653.7&      \nodata& 17.8$\pm$0.1&      \nodata& 16.7$\pm$0.3\\
Phoenix&  SSTU J015046.46$-$442618.3&      \nodata& 18.7$\pm$0.2&      \nodata& 16.0$\pm$0.2\\
Phoenix&  SSTU J015047.39$-$442634.8&      \nodata& 17.0$\pm$0.1&      \nodata& 15.5$\pm$0.1\\
Phoenix&  SSTU J015047.77$-$442624.4&      \nodata& 19.9$\pm$0.5&      \nodata& 16.8$\pm$0.3\\
Phoenix&  SSTU J015049.42$-$442835.3& 16.8$\pm$0.1&      \nodata& 16.3$\pm$0.1&      \nodata\\
Phoenix&  SSTU J015049.43$-$442739.9& 19.7$\pm$0.3& 19.1$\pm$0.2&      \nodata&      \nodata\\
Phoenix&  SSTU J015049.57$-$442551.3&      \nodata& 19.7$\pm$0.4&      \nodata& 17.0$\pm$0.4\\
Phoenix&  SSTU J015049.70$-$442729.4& 19.1$\pm$0.2& 19.2$\pm$0.2&      \nodata&      \nodata\\
Phoenix&  SSTU J015049.85$-$442753.1& 19.4$\pm$0.2& 18.6$\pm$0.3& 17.3$\pm$0.2&      \nodata\\

\enddata
\tablecomments{The full version of this table includes the IRAC photometry for all eight galaxies and is available electronically. The source ID follows the standard {\it Spitzer} naming convention, giving the truncated (J2000) coordinates. }
\end{deluxetable}

\clearpage

\begin{deluxetable}{lcccc}
\tablewidth{0pc}
\tabletypesize{\normalsize}
\tablecolumns{5}
\tablecaption{Nearby background Galaxy Clusters\label{tab:galclust}}

\tablehead{\colhead{dIrr Galaxy} & \colhead{Cluster Name}
  & \colhead{Angular} & \colhead{Angular} & \colhead{Cluster
  Population}\\ & & \colhead{Distance\tablenotemark{a}} &
  \colhead{Cluster Radius} & \colhead{(No. Galaxies)}}

\startdata
LGS~3   & Zwicky 317  & 20\arcmin{} & 16.2\arcmin{} & 64 \\
 & Zwicky 337  & 43\arcmin{} & 9\arcmin{}    & 66 \\
Leo~A   & Zwicky 2776 & 15\arcmin{} & 7\arcmin{}    & 90 \\
 & Zwicky 2778 & 22\arcmin{} & 13\arcmin{}   & 107 \\
 & Zwicky 2827 & 40\arcmin{} & 16\arcmin{}   & 123 \\
 & Zwicky 2813 & 50\arcmin{} & 7\arcmin{}    & 76 \\
 & Zwicky 2866 & 58\arcmin{} & 6\arcmin{}    & 77 \\
Pegasus~dIrr & Zwicky 8949 & 60\arcmin{} & 8\arcmin{}  & 57 \\
 & Zwicky 8933 & 60\arcmin{} & 58\arcmin{} & 352 \\
WLM & CEDAG 12    & 26\arcmin{} & \nodata & 50 -- 80 \\
 & CEDAG 10405 & 26\arcmin{} & \nodata & 80 -- 130 \\
IC~1613 & Zwicky 336 & 25\arcmin{} & 13\arcmin{} & 114 \\
 & CEDAG 461 & 55\arcmin{} & \nodata & 30 -- 40 \\ 
\enddata
\tablecomments{Galaxy clusters within one degree of our targets
  are listed here \citep{abell95, fernandez96}.}
\tablenotetext{a}{ \ Projected angular distance of the dIrr galaxy
  from the center of the background galaxy cluster.}
\end{deluxetable}

\clearpage

\begin{deluxetable}{lcccccccc}
\tablewidth{0pc}
\rotate
\tabletypesize{\small}
\tablecolumns{9} 
\tablecaption{Detection Statistics\label{tab:detstats}}

\tablehead{\multicolumn{9}{c}{Total Point-Source 4\,$\sigma$ Detections in All Wavebands}}

\startdata

\colhead{Filter}& \colhead{Phoenix} &\colhead{LGS~3}&\colhead{DDO~210}
& \colhead{Leo~A} & \colhead{Pegasus~dIrr} & \colhead{Sextans~A} &\colhead{WLM} &\colhead{IC~1613}\\
&\multicolumn{8}{c}{(Number)}\\
\hline
Both {\it V} and {\it I}\dotfill  & 2026 & 369 & 1532 & 2525 &2251 &2968&4896&7574\\
3.6~\micron{}\dotfill & 1087 & 283 & 907 & 778 & 1422 & 1162 &2855&5556\\
4.5~\micron{}\dotfill & 1061 & 285 & 893 & 784 & 1369 & 1204 &2019&3362\\
5.8~\micron{}\dotfill &  462 & 49  & 326 & 261 &  451 &  360 &300&998\\
8.0~\micron{}\dotfill &  107 & 50  & 184 & 148 &  204 &  283 &122&696\\
\hline

\\[-0.3em]
\multicolumn{9}{c}{4\,$\sigma$ Detections in All Four IRAC Bands, But Not {\it V}
  and {\it I}}\\
\hline
\\[-0.7em]
\colhead{Object Type} &\multicolumn{8}{c}{(Number)} \\

\hline
Total\dotfill & 40 & \nodata & 27 & 63 & 67 & 66 &46&190\\
Above AGB limit\dotfill & 1 & \nodata & 3 & 3 & 5 & 11 &3&4\\
Above TRGB, below AGB limit\dotfill & 14 & \nodata & 22 &60 & 58 & 55 &42&183\\
Below TRGB\dotfill & 25 & \nodata & 2 & 0 & 4 & 0 &1&6\\
\hline

\\[-0.7em]
\multicolumn{9}{c}{3.6~\micron{} Point-Source Flux}\\
\hline
\\[-0.7em]

Total Flux (mJy)\dotfill & 19.6& 13.0& 22.9 & 19.6& 23.9& 26.9&40.1&74.0\\
Average 3.6~\micron{} flux (mJy
arcmin$^{-2}$)\dotfill&0.6&0.4&0.7&0.6&0.7&0.8 &1.2&2.2\\
Flux 1~mag Above/Below TRGB\tablenotemark{a}\dotfill & 0.60 & 1.59 & 1.78 & 1.36 & 1.11 &
3.30 & 1.41 & 1.13\\ 
$f_{\rm 1~Gyr}$ \citep{orban08}\tablenotemark{b}\dotfill& 0.027&0.015&0.037 & 0.13 &
0.057 & 0.15 & 0.14 & 0.059\\

\enddata

\tablecomments{Numbers in this table are computed only for areas where
the IRAC and optical coverage overlaps. Some entries for LGS~3 are
left blank due to the difficulty of matching IRAC data with the
high-resolution HST data (\S\,\ref{sec:optphot}). Stars brighter than the {\it I}-band TRGB in
LGS~3 were matched to their IR counterparts by eye. See Papers I and
II for similar tables for WLM and IC 1613.}

\tablenotetext{a}{\ The ratio of the cumulative flux one magnitude
  brighter than (above) to one magnitude fainter than (below) the
  3.6~\micron{}. The uncertainty of this quantity ranges from 5\% to
  18\%.}

\tablenotetext{b}{\ The fraction of the total stellar mass that has
  been formed within the last 1~Gyr \citep{orban08}.}
\end{deluxetable}

\clearpage

\begin{deluxetable}{lcccccccc}
\tablewidth{0pc}
\tabletypesize{\footnotesize}
\rotate
\tablecolumns{9} 
\tablecaption{Detection Statistics of IR-Identified AGB Candidates\label{tab:compfrac}}
\tablehead{& \colhead{Phoenix} &\colhead{LGS~3}&\colhead{DDO~210}
& \colhead{Leo~A} & \colhead{Pegasus~dIrr} & \colhead{Sextans~A} &
  \colhead{WLM\tablenotemark{a}} & \colhead{IC 1613\tablenotemark{a}}}

\startdata

No. IR AGB candidates\tablenotemark{b}\dotfill & 57 & 102 &447 & 302 & 422 & 673&612&956\\
No. foreground stars\tablenotemark{c}\dotfill&6&10&32&6&10&10&9&15\\
No. background galaxies\tablenotemark{d}&20&68&287&180&105&195&161&362\\
No. known carbon stars\tablenotemark{e}&3&0&7&0&32&0&100&70\\
\hline
\\[-0.5em]
\multicolumn{9}{l}{IR AGB Candidates Detected With:}\\
\\[-0.7em]

$M_{\rm I} < -2.5$~mag\dotfill   & 68\% &\nodata&64\%&28\%& 56\%& 55\%&58\%&54\%\\
$M_{\rm I} <$ {\it I}-band TRGB\dotfill & 44\% & 12\%& 40\%&21\%& 52\%&54\%&46\%&49\%\\
\hspace{2em}Corrected\tablenotemark{f}\dotfill &84\%&50\%&140\%&54\%&72\%&78\%&64\%&81\%\\

\hline
\\[-0.5em]
\multicolumn{9}{l}{Stellar Mass Derived from IR AGB Candidates:}\\
\\[-0.7em]

log($M_{2Gyr}^{*}$) ($M_\odot$)\tablenotemark{g}\dotfill & 5.74 & 6.20 & 6.51 & 6.36 & 6.64 & 6.77&6.85&7.07\\
log($M_{SFH}^{*}$) ($M_\odot$)\tablenotemark{h}\dotfill & 5.20 -- 5.74
& 5.75 -- 6.20 & 5.92 -- 6.81 & 5.77 -- 5.91 & 6.19 -- 6.94 & 6.17 --
6.20&6.40 -- 6.85& 6.62 --7.374\\
log($M_*$) ($M_\odot$)\tablenotemark{i}\dotfill & 5.40 & 4.56 & 5.60 & 5.89 & 6.98 & 6.24 & 6.88 & 6.82\\

\enddata

\tablecomments{Numbers in this table are computed only for areas where
the IRAC and optical coverage overlaps. Some entries for LGS~3 are
left blank due to the difficulty of matching IRAC data with the
high-resolution HST data (\S\,\ref{sec:optphot}). Stars brighter than the {\it I}-band TRGB in
LGS~3 were matched to their IR counterparts by eye.}

\tablenotetext{a}{ \ Similar quantities are presented for WLM in Paper
I and for IC~1613 in Paper II. We present updated values here for
these two galaxies using the same analysis procedure used for the six
new galaxies in the sample.}

\tablenotetext{b}{ \ IR-identified AGB candidates are those between
the 3.6~\micron{} TRGB and the AGB limit that are not identified as
either a blue object or an RSG in the optical
(\S\,\ref{sec:agbs}). This quantity includes contamination from
background galaxies and foreground stars, also listed in this Table.}

 \tablenotetext{c}{ \ Estimated number of foreground stars from
\citet{robin03}.}

\tablenotetext{d}{\ Estimated number of unresolved background galaxies
above the 3.6~\micron{} TRGB found through fitting the radial profiles
of AGB candidates (see \S\,\ref{sec:bkgnd}). These estimates are
within 25\% of the estimates measured from S-COSMOS data for four
galaxies, 40\% for three, and 78\% for IC~1613.}

\tablenotetext{e}{ \ Carbon stars listed are those detected in
  IRAC. The carbon star searches were carried out by:
  \citet{battinelli00} for DDO~210 and Pegasus~dIrr, \citet{dacosta94}
  and \citet{menzies08} for Phoenix, \citet{battinelli04} for WLM, and
  \citet{albert00} for IC~1613.}

\tablenotetext{f}{\ IR AGB candidates detected with $M_{\rm I} <
  I$-band TRGB, with subtraction of foreground stars and unresolved
  background galaxies from radial profile fitting (see
  \S\,\ref{sec:bkgnd}).}

\tablenotetext{g}{ \ Total stellar mass ($M_{2Gyr}^{*}$) was
determined following \citet{vanloon05}, assuming an age of 2~Gyr, and
using the number of 3.6~\micron{} stars above the TRGB, excluding
estimates of the numbers of background galaxies from S-COSMOS and
foreground stars from \citet{robin03}.}

\tablenotetext{h}{ \ Total stellar mass ($M_{SFH}^{*}$) was determined
following \citet{vanloon05}, assuming a single age range based on past
star formation events \citep{dolphin05}, and using the number of
3.6~\micron{} stars above the TRGB, excluding estimates of the numbers
of background galaxies from S-COSMOS and foreground stars from
\citet{robin03}.}

\tablenotetext{I}{ \ Total stellar mass from Table~\ref{tab:properties}.}

\end{deluxetable}

\clearpage

\begin{deluxetable}{lccccc}
\tablewidth{0pc}
\tabletypesize{\normalsize}
\tablecolumns{6} 
\tablecaption{Average Stellar Mass-Loss Rates\label{tab:avgmlr}}

\tablehead{ \multicolumn{6}{c}{All IR Sources With $M_{\rm 3.6} <$ TRGB
    ($10^{-6}~M_{\odot}~yr^{-1}$)}\\[-0.5em]\\\colhead{Galaxy}&\colhead{C-rich 1\tablenotemark{a}} & \colhead{C-rich 2\tablenotemark{b}} & \colhead{O-rich 1\tablenotemark{c}} &\colhead{O-rich 2\tablenotemark{d}}&\colhead{O-rich 3\tablenotemark{e}}}

\startdata

Phoenix   & $0.53$ & $0.31$ & $1.4$ & $1.2$ & $1.1$\\
LGS~3     & $2.21$& $1.4$ & $5.4$ & $4.6$ & $4.6$\\
DDO~210   & $2.7$ & $1.4$ & $5.8$ & $4.9$ & $4.6$\\
Leo~A     & $1.6$ & $1.2$ & $4.0$ & $3.4$ & $3.7$\\
Pegasus~dIrr   & $2.1$ & $0.88$ & $4.4$ & $3.9$ & $2.6$\\
Sextans~A & $2.4$ & $1.2$ & $5.4$ & $4.5$ & $3.6$\\
WLM       & $3.5$ & $1.1$ & $6.6$ & $5.7$ & $3.5$\\
IC~1613   & $1.1$ & $0.47$ & $2.1$ & $1.8$ & $1.5$\\
\enddata
\tablecomments{The mass-loss rates quoted here include only sources located between
the 3.6~\micron{} TRGB and the AGB limit that are not optically
identified as a blue object or RSG. Due to color limits in the
\citet{groenewegen06} models, C-rich~2 and O-rich~1 include the
most sources, while C-rich~1 includes the fewest.}
\tablenotetext{a}{ \ C-rich AGB star, $T_{\rm eff} = 2650$ K, dust
  composition is 85\% amorphous carbon, 15\% SiC.}
\tablenotetext{b}{ \ C-rich AGB star, $T_{\rm eff} = 3600$ K, dust
  composition is 85\% amorphous carbon, 15\% SiC.}
\tablenotetext{c}{ \ O-rich AGB star, $T_{\rm eff} = 2500$ K, dust
  composition is 60\% silicates, 40\% aluminum oxides.}
\tablenotetext{d}{ \ O-rich AGB star, $T_{\rm eff} = 2500$ K, dust
  composition is 100\% silicates.}
\tablenotetext{e}{ \ O-rich AGB star, $T_{\rm eff} = 3300$ K, dust
  composition is 60\% silicates, 40\% aluminum oxides.}

\end{deluxetable}

\clearpage

\begin{deluxetable}{lccccc}
\tablewidth{0pc}
\tabletypesize{\footnotesize}
\tablecolumns{6} \tablecaption{Integrated Galaxy Mass-Loss
Rates\label{tab:cummlr}}

\tablehead{ \multicolumn{6}{c}{All AGB Candidates ($M_{\rm 3.6} <$
    TRGB; $10^{-4}~M_{\odot}~yr^{-1}$)}\\[-0.5em]\\\colhead{Galaxy}&\colhead{C-rich
    1\tablenotemark{a}} & \colhead{C-rich 2\tablenotemark{a}} &
    \colhead{O-rich 1\tablenotemark{a}} &\colhead{O-rich
    2\tablenotemark{a}}&\colhead{O-rich 3\tablenotemark{a}}}

\startdata
Phoenix   & $0.05$ & $0.18$ & $0.29$& $0.25$& $0.65$\\
LGS~3     & $2.4$ & $3.7$ & $9.5$ & $7.9$ & $12$\\
DDO~210   & $39$ & $39$ & $75$ & $94$ & $76$\\
Leo~A     & $2.0$ & $3.3$ & $8.6$ & $7.3$ & $11$\\
Pegasus~dIrr   & $4.1$ & $4.7$ & $12$ & $12$ & $13$\\
Sextans~A & $5.0$ & $6.9$ & $19$ & $16$ & $23$\\
WLM       & $7.8$ & $9.2$ & $25$ & $23$ & $28$\\
IC~1613   & $2.4$ & $4.2$ & $10$ & $8.8$ & $14$\\
\hline
\\[-0.5em]

\multicolumn{6}{c}{All AGB Candidates With $M_{\rm I} < -2.5$ ($10^{-4}~M_{\odot}~yr^{-1}$)}\\ \hline

Phoenix   & $0.03$ & $0.11$ & $0.15$ & $0.13$ & $0.20$\\
LGS~3     & \nodata & \nodata & \nodata & \nodata & \nodata\\
DDO~210   & $1.1$ & $2.3$ & $5.0$ & $4.3$ & $8.0$\\
Leo~A     & $0.57$ & $0.82$& $2.0$ & $1.6$ & $2.6$\\
Pegasus~dIrr   & $3.3$ & $3.7$ & $9.3$ & $9.2$ & $10$\\
Sextans~A & $1.6$ & $2.8$ & $6.9$ & $5.7$ & $9.5$\\
WLM       & $2.2$ & $2.8$ & $8.0$ & $6.9$ & $9.5$\\
IC~1613   & $0.67$ & $1.6$ & $3.4$& $2.9$& $5.7$\\
\hline
\\[-0.5em]

\multicolumn{6}{c}{All AGB Candidates With $M_{\rm I} < -2.5$}\\
\multicolumn{6}{c}{Or With $m_{\rm 3.6} <$ 16 ($10^{-4}~M_{\odot}~yr^{-1}$)}\\
\hline

Phoenix   & $0.03$ & $0.12$ & $0.15$ & $0.13$ & $0.44$\\
LGS~3     & \nodata & \nodata & \nodata & \nodata & \nodata\\
DDO~210   & $1.5$ & $2.7$ & $6.1$ & $5.2$ & $9.2$\\
Leo~A     & $0.94$& $1.5$ & $3.7$ & $3.1$ & $4.8$\\
Pegasus~dIrr   & $3.3$ & $3.8$ & $9.4$ & $9.4$ & $10$\\
Sextans~A & $1.6$ & $2.8$ & $7.1$ & $5.9$ & $9.8$\\
WLM       & $3.4$ & $4.2$ & $12$ & $10$ & $14$\\
IC~1613   & $1.2$ & $2.3$ & $5.3$ & $4.5$ & $8.0$\\
\hline
\\[-0.5em]

\multicolumn{6}{c}{All AGB Candidates Minus the}\\
\multicolumn{6}{c}{No. Galaxies Times the Average Stellar MLR
($10^{-4}~M_{\odot}~yr^{-1}$)\tablenotemark{b}}\\ \hline

Phoenix   & $0.0$ & $0.11$ & $0.04$ & $0.03$ & $0.42$\\
LGS~3     & $0.11$ & $1.4$ & $0.84$ & $0.71$ & $4.7$\\
DDO~210   & $0.25$ & $3.2$ & $4.0$ & $3.4$ & $11 $\\
Leo~A     & $0.38$ & $1.5$ & $2.5$ & $2.1$ & $5.1$\\
Pegasus~dIrr   & $0.38$ & $1.8$ & $2.1$ & $1.8$ & $5.9$\\
Sextans~A & $0.0 $ & $2.5$ & $0.13$ & $0.11$ & $9.2$\\
WLM       & $1.2$ & $4.3$ & $6.6$ & $5.7$ & $15$\\
IC~1613   & $0.64$ & $2.8$ & $4.8$ & $4.0$ & $9.6$\\

\enddata
\tablecomments{The mass-loss rates quoted here include only stars located between
the 3.6~\micron{} TRGB and the AGB limit that are not optically
identified as a blue object or RSG.Due to color limits in the
\citet{groenewegen06} models, C-rich~2 and O-rich~1 include the
most sources, while C-rich~1 includes the fewest.}
\tablenotetext{a}{ See Table~\ref{tab:avgmlr} for descriptions of
  these labels.}
\tablenotetext{b}{ The number of galaxies here were estimated from S-Cosmos data.}
\end{deluxetable}

\clearpage

\begin{deluxetable}{lcccccccc}
\tablewidth{0pc}
\tabletypesize{\small}
\tablecolumns{9} \tablecaption{ISM Gas Return\label{tab:ism}}

\tablehead{\colhead{Galaxy} & 
  \colhead{$\dot{M}_{\rm dusty}$\tablenotemark{a}} &
  \colhead{${\rm SFR}_{\rm (1\,Gyr)}$\tablenotemark{b}} &
  \colhead{${\rm SFR}_{\rm (5\,Gyr)}$\tablenotemark{b}} &
  \colhead{$f_{\rm return}$\tablenotemark{c}} & 
  \colhead{$\tau_{\rm R}$\tablenotemark{d}} & 
  \colhead{$\tau_{\rm R, corrected}$\tablenotemark{d}} & 
  \colhead{$\tau_{\rm lifetime}$\tablenotemark{e}} &
  \colhead{$\tau_{\rm age}$\tablenotemark{b}} \\[0.5em]
  &
  \colhead{($M_\odot~{\rm yr}^{-1}$)} &
  \colhead{($M_\odot~{\rm yr}^{-1}$)} &
  \colhead{($M_\odot~{\rm yr}^{-1}$)} &
  &
  \colhead{(Gyr)} &
  \colhead{(Gyr)} &
  \colhead{(Gyr)} &
  \colhead{(Gyr)} }

\startdata

Phoenix      &$1.2 \times 10^{-5}$&$5.4 \times 10^{-5}$&$9.2 \times 10^{-5}$&0.22&59 &76&43&10.3\\
LGS~3        &$1.6 \times 10^{-4}$&$3.0 \times 10^{-5}$&$6.4 \times 10^{-5}$&5.30&3.3&\nodata&\nodata&9.8\\
DDO~210      &$4.4 \times 10^{-4}$&$1.9 \times 10^{-4}$&$1.2 \times 10^{-4}$&2.32&1.1&\nodata&\nodata&12.0\\
Leo~A        &$2.3 \times 10^{-4}$&$8.2 \times 10^{-4}$&$8.2 \times 10^{-4}$&0.24&12&17&18&6.2\\
Pegasus~dIrr &$2.4 \times 10^{-4}$&$8.0 \times 10^{-4}$&$1.3 \times 10^{-3}$&0.30&3.5&4.8&12&7.4\\
Sextans~A    &$2.4 \times 10^{-4}$&$1.2 \times 10^{-2}$&$6.0 \times 10^{-3}$&0.02&6.6&6.8&14&9.3\\
WLM          &$6.6 \times 10^{-4}$&$2.2 \times 10^{-2}$&$1.7 \times 10^{-2}$&0.03&2.4&2.5&3.7&6.7\\
IC~1613      &$4.4 \times 10^{-4}$&$1.2 \times 10^{-2}$&$1.7 \times 10^{-2}$&0.04&5.5&5.7&4.3&7.7\\

\enddata 

\tablecomments{ An empty entry indicates an indefinite
    timescale. }

\tablenotetext{a}{ $\dot{M}_{\rm dusty}$ is the MLR from the bottom
    panel of Table~\ref{tab:cummlr}, averaged over all compositions.}

\tablenotetext{b}{ Average SFRs and the mean mass-weighted stellar
    age ($\tau_{\rm age}$) are from \citet{orban08}.}

\tablenotetext{c}{ $f_{\rm return} = \dot{M}_{\rm
    dusty}~/~{\rm SFR}_{\rm (1\,Gyr)}$.}

\tablenotetext{d}{ $\tau_{\rm R}$ is the Roberts Time: $\tau_{\rm R} =
    M_{\rm gas}~/~{\rm SFR}_{\rm (1\,Gyr)}$. The correction
    factor (recycling factor) for $\tau_{\rm R}$ is $(1 - \dot{M}_{\rm
    dusty}~/~{\rm SFR}_{\rm (1\,Gyr)})^{-1}$.}

\tablenotetext{e}{ $\tau_{\rm lifetime}$ is the expected total
  gas-rich lifetime of a galaxy assuming a total initial mass that
  includes the stellar and \ion{H}{1} mass, that the current dusty MLR
  has remained constant, and that the 5~Gyr average SFR represents the
  lifetime average SFR: $\tau_{\rm lifetime} = M_{\rm
  Total}~/~({\rm SFR}_{(\rm 5\,Gyr)} - \dot{M}_{\rm dusty})$. }

\end{deluxetable}

\end{document}